\documentclass[useAMS,usenatbib,a4paper]{mn2e_mod}
\usepackage{epsfig,amsmath,amssymb,amsfonts,natbib,color,url}  

\usepackage{amsmath,amssymb,amsfonts,natbib}
\usepackage {graphicx,color}
\usepackage {txfonts }
\usepackage {natbib  }
\usepackage {float   }
\usepackage {enumerate}
\usepackage {subfigure  }

\newcommand{\revised}[1]{{#1}}

\title[Colliding Planetary and Stellar Winds]{Colliding Planetary and Stellar Winds: \\
Charge Exchange and Transit Spectroscopy in Neutral Hydrogen}
\author[P.~Tremblin \&\,E.\,Chiang]{Pascal Tremblin$^{1}$\footnotemark[1] and\,Eugene~Chiang$^{2}$\\
 \\
  $^1$Laboratoire AIM Paris-Saclay (CEA/Irfu - Uni. Paris Diderot - CNRS/INSU), 
           Centre d'\'etudes de Saclay,  91191 Gif-Sur-Yvette, France\\
  $^2$Departments of Astronomy and of Earth and Planetary Science, University of California at Berkeley, Hearst Field Annex B-20, Berkeley CA 94720-3411, USA\\}

\date{Submitted: \today}

\begin{document}
\maketitle
\begin{abstract}
  When transiting their host stars, hot Jupiters absorb about 10\% of
  the light in the wings of the stellar Lyman-alpha emission
  line. \revised{
The absorption occurs at wavelengths
  Doppler-shifted from line center by $\pm 100$ km/s --- larger
  than the thermal speeds with which partially neutral, 
  $\sim$$10^4$ K hydrogen
  escapes from hot Jupiter atmospheres.}
It has been proposed that the
  absorption arises from $\sim$$10^6$ K hydrogen from the host stellar wind,
  made momentarily neutral by charge exchange with planetary H I. 
  \revised{
The $\pm$100 km/s velocities would then be attributed to the typical
velocity dispersions of protons in the stellar
wind --- as inferred from spacecraft measurements of the Solar wind.
}
To test this proposal, we perform 2D hydrodynamic simulations of
  colliding hot Jupiter and stellar winds, augmented by a chemistry
  module to compute the amount of hot neutral hydrogen produced by
  charge exchange. We observe the contact discontinuity where the two
  winds meet to be Kelvin-Helmholtz unstable. The Kelvin-Helmholtz
  instability mixes the two winds; in the mixing layer, charge
  exchange reactions establish, within tens of seconds, a chemical
  equilibrium in which the neutral fraction of hot stellar hydrogen
  equals the neutral fraction of cold planetary hydrogen (about
  20\%). In our simulations, enough hot neutral hydrogen is generated
  to reproduce the transit observations, and the amount of absorption
  converges with both spatial resolution and time. 
%
%
 Our calculations support the idea that charge
  transfer between colliding winds correctly explains the Lyman-alpha
  transit observations --- \revised{
modulo the effects of magnetic fields, which we do not model
but which may suppress mixing.
Other neglected effects include, in order of decreasing
importance, rotational forces related to orbital motion, gravity,
and stellar radiation pressure; we discuss quantitatively
the errors introduced by our
approximations.}
  How hot stellar hydrogen cools when it collides with cold
  planetary hydrogen is also considered; a more careful treatment
  of how the mixing layer thermally equilibrates might explain the
  recent detection of Balmer H$\alpha$ absorption in transiting hot
  Jupiters.
\end{abstract}

\label{firstpage}
\footnotetext[1]{e-mail: \rm{\url{pascal.tremblin@cea.fr}.}}

%
%

\section{INTRODUCTION}\label{sec:intro}

Gas-laden planets lose mass to space when their upper atmospheres are
heated by stellar ultraviolet (UV) radiation.  Ubiquitous in the Solar
System, thermally-driven outflows modify the compositions of their
underlying atmospheres over geologic time (e.g.,
\citealt{weissman}). Thanks to the {\it Hubble Space Telescope (HST)},
escaping winds are now observed from extrasolar hot Jupiters:
Jovian-sized planets orbiting at distances $\lesssim 0.05$ AU from
their host stars and bathed in intense ionizing fields. Spectroscopy
with {\it HST} reveals absorption depths of $\sim$2--10\% in various
resonance transitions (H I, O I, C II, Si III and Mg II) when the
planet transits the star, implying gas outflows that extend for at
least several planetary radii (e.g., \citealt{Vidal-Madjar2003}, VM03;
\citealt{Vidal-Madjar2004}; \citealt{Ben-Jaffel2007};
\citealt{Ben-Jaffel2008}; \citealt{Vidal-Madjar2008};
\citealt{Lecavelier2010}; \citealt{Fossati2010};
\citealt{Linsky2010}). 
\revised{
Recent observations of HD 189733b also
indicate temporal variations in H I Lyman-$\alpha$
absorption, possibly correlated with stellar X-ray
activity \citep[][]{LecavelierdesEtangs:2012jq}.} 
These data promise to constrain the compositions
of hot Jupiter atmospheres and the degrees to which they are vertically mixed
(\citealt{Liang2003}; \citealt{Moses2011}). 

The {\it HST} observations of hot Jupiter winds are accompanied by
theoretical studies that model planetary outflows starting from first
principles (e.g., \citealt{Yelle2004}; 
\revised{
\citealt{Yelle:2006fw}; 
}
\citealt{Tian2005};
\citealt{GarciaMunoz2007}; Murray-Clay, Chiang, \& Murray~2009, M09).
These 1D hydrodynamic models generally agree that hot Jupiters like HD
209458b and HD 189733b are emitting $\dot{M} \sim 10^{10}$--$10^{11}$
g/s in mostly hydrogen gas. 
\revised{
Three-dimensional models include \citet{LecavelierdesEtangs:2004jm} and
\citet{Jaritz:2005kr}, who emphasize the importance of tidal forces.
}

Do the models agree with the observations?  Linsky et al.~(2010) find
that their observations of C II absorption in HD 209458b can be made
consistent with modeled mass loss rates, assuming the carbon abundance
of the wind is not too different from solar. More comparisons between
observation and theory would be welcome---particularly for hydrogen,
the dominant component of the wind. But the observations of H I
absorption have proven surprisingly difficult to interpret. On the one
hand, the original measurements by VM03 indicate substantial
($\sim$10\%) absorption at Doppler shifts of $\pm 100$ km/s from the
center of the H I Lyman-$\alpha$ line. On the other hand, theory
(e.g., M09) indicates that planetary outflows, heated by
photoionization to temperatures $T \lesssim 10^4$ K, blow only at
$\sim$10 km/s.  How can such slow planetary winds produce significant
absorption at $\pm 100$ km/s? \revised{
Measurements of blueshifted velocities
as large as -230 km/s in the case of HD 189733b only accentuate
this problem \citep{LecavelierdesEtangs:2012jq}.
}

\citet[][H08]{Holmstrom2008} 
propose that the observed energetic neutral H atoms arise from charge
exchange between planetary H I and protons from the incident {\it
  stellar} wind. In this interpretation, the $\pm$100 km/s velocities
correspond to the thermal velocities of $10^6$ K hydrogen from the
star---hydrogen which is made neutral by electron-exchange with planetary H I.
The situation is analogous to that of the colliding winds of O star
binaries (\citealt{Stevens:1992gf}; \citealt{lamberts}, and references
therein). The H I Lyman-$\alpha$ absorption arises from the contact
discontinuity where the two winds meet, mix, and charge exchange to
produce hot neutral hydrogen. 

The calculations of H08, and those of the follow-up study by
\citet[][E10]{Ekenback2010}, are based on a Monte Carlo algorithm
that tracks individual ``meta-particles'' of neutral hydrogen launched
from the planet. The meta-particles collide and charge exchange with
stellar wind protons outside a presumed planetary magnetosphere, which
is modeled as an ``obstacle'' in the shape of a bow shock. Good
agreement with the Ly-$\alpha$ observations is obtained for a range of
stellar and planetary wind parameters, and for a range of assumed
obstacle sizes.

In this work we further test the hypothesis of charge exchange first
explored by H08 and E10.  Our methods are complementary: instead of
adopting their kinetic approach, we solve the hydrodynamic equations.
We do not prescribe any obstacle to deflect the stellar wind, but
instead allow the planetary and stellar winds to meet and shape each
other self-consistently via their ram and thermal pressures. Some
aspects of our solution are not realistic---we ignore the Coriolis
force, the centrifugal force, stellar tidal gravity, and magnetic
fields.\footnote{
\revised{
For recent explorations of star-planet interactions
including magnetic forces, see \citet{cohenetal11a,cohenetal11b}.
These simulations do not resolve the mixing layer interface between the stellar
and planetary winds.}}
Our goal is to develop a first-cut hydrodynamic-chemical model of the contact
discontinuity between the two winds where material mixes and charge
exchanges. Simple and physically motivated scaling relations
will be developed between the amount of H I absorption and the
properties of the stellar and planetary winds.

The plan of this paper is as follows. In \S\ref{sec:method} we
describe our numerical methods, which involve augmenting our
grid-based hydrodynamics code to solve the chemical reactions of
charge exchange, and specifying special boundary conditions to launch
the two winds.  In \S\ref{sec:results} we present our results,
including a direct comparison with the H I Ly$-\alpha$ transit spectra
of VM03, and a parameter study to elucidate how the absorption depth
varies with stellar and planetary wind properties. A summary is given
in \S\ref{sec:summary}, together with an assessment of the
shortcomings of our study and pointers toward future work.

%
%

\section{NUMERICAL METHODS} \label{sec:method}

In \S\ref{sec:hydro} we describe the hydrodynamics code
used to simulate the colliding planetary and stellar
winds. In \S\ref{sec:ce} we detail the charge exchange reactions
that were added to the code. In \S\ref{sec:lyman},
we outline our post-processing procedure for
computing the Lyman-$\alpha$ transmission spectrum.
As a convenience to readers, in \S\ref{sec:difference} we re-cap
the differences between our treatment of colliding winds
and that of E10/H08.

\subsection{Hydrodynamics: Code and Initial Conditions} \label{sec:hydro}

Our simulations are performed with \texttt{HERACLES} \citep{Gonzalez2007},\footnote{http://irfu.cea.fr/Projets/Site\_heracles/index.html} a grid-based code using a second-order Godunov scheme to solve the Euler equations: 
\begin{eqnarray}\label{euler}
\frac{\partial \rho}           {\partial t} + \nabla \cdot (\rho \textbf{V})                         & = & 0 \cr
\frac{\partial \rho \textbf{V}}{\partial t} + \nabla \cdot [\rho \textbf{V} \otimes \textbf{V} + pI] & = & 0 \cr
\frac{\partial E}              {\partial t} + \nabla \cdot [(E + p) \textbf{V}                     ] & = & 0 \cr
\frac{\partial \rho x_i}           {\partial t} + \nabla \cdot (\rho x_i\textbf{V})                         & = & 0 \quad.
\end {eqnarray}
Here $\rho$, $\textbf{V}$, $p$, and $E$ are the mass density, velocity,
pressure, and total energy density, respectively (e.g., \citealt{Clarke2003}).
The code tracks abundances of individual species:
$x_i$ is the mass fraction of the $i^{\rm th}$ species of hydrogen, where
$i \in \{1,2,3,4\}$ to cover four possible combinations of ionization
state (either neutral or ionized) and temperature (either ``hot'' because it
is arises from the star or ``cold'' because it arises from the planet).
The outer product is denoted $\otimes$, and $I$ is the identity matrix.

All our simulations are 2D Cartesian in the dimensions $x$
(stellocentric radius) and $y$ (height above the planet's orbital
plane). Equivalently the simulations may be regarded as 3D, but with
no rotation and with a star and a planet that are infinite cylinders oriented
parallel to the $z$-axis. At fixed computational cost,
two-dimensional simulations enjoy better spatial resolution
than three-dimensional simulations and thus
better resolve the fluid instabilities at the interface
of the two winds. The standard box size is $(L_x, L_y) = (40 R_p, 60 R_p)$, where $R_p = 10^{10}$ cm is the planet radius. The number
of grid points ranges up to $(N_x, N_y) = (6400,9600)$; see Table 1.

The star and its wind are modeled after the Sun and the Solar wind.
The stellar wind is injected through the left edge of the simulation
box; the densities, velocities, and temperatures in the
vertical column of cells at the box's left edge are fixed in time. 
Stellar wind properties as listed in Table \ref{parameters} are
given for a stellocentric distance $r = r_{{\rm launch},\ast} = 5R_\odot$,
near the box's left edge. Here the stellar wind
density, temperature, sound speed,
and flow speed are set to $n_\ast = 2.9 \times 10^4$ cm$^{-3}$,
$T_\ast = 10^6$ K, $c_\ast = 129$ km/s (computed for a mean molecular weight equal
to half the proton mass, appropriate for an $f_\ast^+$ = 100\% ionized
hydrogen plasma), and $v_\ast = 130$ km/s
(\citealt{Sheeley1997}; \citealt{quemerais2007}; see also \citealt{lemaire2011}),
respectively. Our stellar wind
parameters are such that the implied spherically symmetric (3D) mass loss
rate is $1 \times 10^{12}$ g/s or $2 \times 10^{-14} M_{\odot}$/yr.

Our stellar wind parameters are similar to those of the ``slow'' Solar wind
in the Sun's equatorial plane. Compare our choices with those
of~Ekenb\"ack et al.~(2011),
who adopt a stellar wind speed of 450 km/s. Their 
speed is closer to that of the ``fast'' Solar wind which emerges from
coronal holes. At Solar minimum, the fast wind tends to be confined to large
heliographic latitudes (polar regions), but at Solar maximum, the coronal
holes migrate to lower latitudes and the fast wind can more readily
penetrate to the ecliptic
(\citealt{kohl98}; \citealt{mccomas2003}; S.~Bale 2012, personal communication).
Evaporating hot Jupiters like HD 209458b and HD 189733b have orbit normals
that are nearly aligned with the spin axes of their host stars
(\citealt{winn2005}; \citealt{winn2006}). Because such planets
reside near their stellar equatorial planes, the slow equatorial Solar wind
seems a better guide than the fast, more polar wind; nevertheless,
as noted above, the fast
wind is known to extend to low latitudes, and the speeds and densities
of both winds vary by factors of order-unity or more with time.

The stellar wind velocity at the left
boundary is not plane-parallel but points radially away from the
central star (located outside the box). The 
density, velocity, and temperature in each cell at
the boundary are computed by assuming
that the central star emits a spherical isothermal wind
whose velocity grows linearly with stellocentric distance $r$ and
whose density decreases as $1/r^3$. These scalings,
which are modeled after empirical Solar wind measurements (e.g., Sheeley et al.~1997) and which maintain constant mass loss rate with $r$,
are used only to define the left-edge boundary conditions
and are not used in the simulation domain.
Outflow boundary conditions are applied at the top, bottom, and right
edges of the box. 

As a final comment about our choice of stellar wind parameters, we 
note that they are valid for the left-edge boundary at
$r = r_{{\rm launch},\ast} = 5 R_\odot$---not for the planet's orbital radius of
$r = 10 R_\odot$. The left-edge boundary must be
far enough away from the planet that the
stellar wind properties at the boundary
are well-approximated by their ``free-stream''
values in the absence of any planetary obstacle. We will see in
\S\ref{sec:results} that the stellar wind slows considerably between
$r = 5R_\odot$ and $r = 10R_{\odot}$ as a consequence
of the oncoming planetary wind. This region of deceleration is absent
from the models of H08 and E10.

A circle of radius $d_{\rm launch,p} = 4R_{\rm p}$, centered
at position $(l_x,l_y) = (30R_{\rm p},30R_{\rm p})$ (where the origin
is located at the bottom left corner of the domain),
defines the boundary where the assumed
isotropic and radial planetary wind is launched. 
The properties
of our simulated planetary wind, which are similar 
to those of the standard supersonic models of HD209458b by
Garc\'ia-Mu\~noz (2007) and \citet{Murray-Clay2009},
are listed in Table
\ref{parameters}, and are constant in time along the circular
boundary. The density and velocity of the planetary wind at this
boundary
are such that if the wind were spherically symmetric, the mass loss rate would be $1.6 \times 10^{11}$ g/s. This value
lies within the range estimated from observations by Linsky et
al.~(2010) 
\revised{
and from energetic considerations 
\citep[e.g.,][]{LecavelierdesEtangs:2007fa,Ehrenreich:2011cp}.
}
Note that $1-f_p^+ = 20$\% of the planetary wind at launch is
neutral \citep{Murray-Clay2009} and available for charge exchange.
\revised{
This neutral fraction represents a balance between photoionizations
by extreme UV radiation and gas advection of neutrals at a planetary altitude
of 4--5 $R_{\rm p}$ \citep{Murray-Clay2009}.
}
The planetary and stellar winds are barely supersonic at launch
(Mach numbers $M_p = 1.2$ and $M_\ast = 1.01$). 

Gravity is neglected, as are all rotational forces. The pressure
$p$ is related to the internal energy density $e = E - \rho V^2/2$
via $p = (\gamma - 1)e$, where $\gamma = 1.01$. That is, gas is
assumed to behave nearly isothermally.
This isothermal assumption should not be taken to mean that the
temperature is the same across the simulation domain; 
the temperature of the stellar wind at injection is $T_\ast = 10^6$ K,
while that of the planetary wind is
$T_p = 7000$ K.\footnote{\revised{
In the standard model of M09, the temperature
starts at $\sim$$10^4$ K 
at a planetocentric radius of $1.1 R_p$ --- consistent
with the observations by \citet{ballester07} ---
and cools to $\sim$3000 K at $4 R_p$.
The temperatures calculated by \citet{GarciaMunoz2007}
at $4 R_p$ are 6000--7000 K.
}}
Rather, the two winds, as long as they remain unmixed, tend to maintain their respective temperatures as they rarefy and compress. In reality,
the stellar wind can keep, in and of itself, a near-isothermal profile on length scales of interest to us because thermal conduction times (estimated, e.g., using the Spitzer conductivity) are short compared to dynamical times. Treating the planetary wind
as an isothermal flow is less well justified, as cooling by adiabatic expansion can be a significant portion of the energy budget  (Garc\'ia-Mu\~noz 2007; M09). Nevertheless the error incurred by assuming the planetary wind is isothermal is small for our standard model
because the planetary wind hardly travels beyond its launch radius of $4R_p$ before it encounters a shock; thus
rarefaction factors are small. Furthermore, as noted above,
shock compression factors are modest because the speed of the planetary wind is only marginally supersonic.
Where the stellar and planetary winds meet
and mix, the code ascribes an intermediate temperature $10^4 \, {\rm   K} < T < 10^6 \, {\rm K}$. This temperature, as computed
by \texttt{HERACLES}, is used only for the hydrodynamic evolution; it is not used for computing either the charge exchange reactions (\S\ref{sec:ce}) or the transmission spectrum (\S\ref{sec:lyman}).

Each simulation is performed in two steps. First only the planetary
wind is launched from its boundary and allowed to fill the entire
domain for $2\times 10^5$ seconds. Second the stellar wind is injected
through the left side of the box, by suitable assignment of ghost
cells. This two-step procedure was found to minimize transients. The
simulations typically run for $2\times 10^6$ s, which corresponds to
$\sim$60 box-crossing times for the stellar wind in the horizontal
direction.

\begin{table}
\caption{\label{parameters} Parameters of the winds at launch, and of
  the simulation box}
\centering
\begin{tabular}{l|l}
\hline
\hline     
Stellar Wind          & Planetary Wind\\
\hline

$r_{{\rm launch},\ast} = 5R_{\odot}$ & $d_{{\rm launch},p} = 4R_p$ \\
$n_\ast$   = 2.9E4/cm$^3$ & $n_p$   = 3.9E6/cm$^3$ \\
$T_\ast$   = 1E6 K      & $T_p$   = 7000 K     \\
$v_\ast$   = 130 km/s   & $v_p$   = 12 km/s    \\
$f_\ast^+$ = 1          & $f_p^+$ = 0.8        \\
$c_\ast$   = 129 km/s   & $c_p$   = 10 km/s    \\
$M_\ast$   = 1.01       & $M_p$   = 1.2        \\
\hline
\hline
radial ($x$) direction & vertical ($y$) direction \\
\hline
L$_x$/R$_p$ = 40  & L$_y$/R$_p$ =  60  \\
N$_x$       = 50, 100, 200,400, & N$_y$       = 75, 150, 300, 600,  \\ 
800, 1600, 3200, 6400 & 1200, 2400, 4800, 9600

\end{tabular}
\end{table}

\subsection{Charge Exchange}\label{sec:ce}

Charge exchange consists of the following forward and reverse reactions:
\begin{eqnarray}\label{ce}
H_h^+ + H_c^0 &\rightleftarrows& H_h^0 + H_c^+
\end{eqnarray}
Hot (subscript $h$) ionized (superscript $+$) hydrogen emitted by the star can collide with cold (subscript $c$) neutral (superscript $0$)
hydrogen emitted by the planet, neutralizing the former and ionizing
the latter while preserving their kinetic energies.
The reverse reaction occurs with an identical
rate coefficient $\beta$ (units of cm$^3$/s; $\beta$ is the cross section multiplied by the relative velocity).


We have added reaction (\ref{ce}) to \texttt{HERACLES} by integrating the
following equations in every grid cell (we refer to this portion of the calculation as the ``chemistry step''):
\begin{align}
\frac{d (n_H x_h^+) }{d t} & =  \beta n^2_H(x_h^0x_c^+-x_h^+x_c^0) \nonumber\\ 
\frac{d (n_H x_c^0) }{d t} & = + \frac{d (n_H x_h^+)}{dt} \nonumber\\
\frac{d (n_H x_c^+) }{d t} & =  -\frac{d (n_H x_h^+)}{dt} \nonumber\\
\frac{d (n_H x_h^0) }{d t} & =  -\frac{d (n_H x_h^+)}{dt} \nonumber\\
                x_h^+ + x_h^0 + x_c^+ + x_c^0 &=  1  \,.
\label{charge_exchange}
\end{align}
Here $n_H$ is the total hydrogen number density (regardless of
ionization state or temperature), and $x_{(c,h)}^{(0,+)}$ is a number fraction
(equivalently a mass fraction because the only element treated in the
simulation is hydrogen). The rate coefficient $\beta = 4 \times
10^{-8}$ cm$^3$/s is calculated by combining the energy-dependent
cross section of \citet{Lindsay2005} with a Maxwellian distribution
for the relative velocity between hydrogen atoms at the two (constant)
temperatures $T_\ast$ and $T_p$.  The finite-difference forms of
equations (\ref{charge_exchange})
are 
\begin{align}
x_h^{+(n+1)} - x_h^{+(n)} = b \left( x_h^{0(n+1)}x_c^{+(n+1)} - x_h^{+(n+1)}x_c^{0(n+1)} \right) \nonumber \\
x_c^{0(n+1)}-x_c^{0(n)} = x_h^{+(n+1)}-x_h^{+(n)} \nonumber \\
x_c^{+(n+1)}-x_c^{+(n)} = -x_h^{+(n+1)} + x_h^{+(n)} \nonumber \\
x_h^{0(n+1)}-x_h^{0(n)} = -x_h^{+(n+1)} + x_h^{+(n)} \nonumber \\
                x_h^{+(n)} + x_h^{0(n)} + x_c^{+(n)} + x_c^{0(n)} &= 1   
\label{implicit}
\end{align}
where the superscript $(n)$ refers to the $n^{\rm th}$ time step, $b
\equiv \beta n_H \Delta t$, and $\Delta t$ is the integration time
step of \texttt{HERACLES}.  Because the righthand side of the first
of these equations is evaluated at step $(n+1)$ instead of step $(n)$, our
scheme is implicit.  The first equation combines with the others to yield
\begin{equation}
x_h^{+(n+1)} = \frac{ \left[ x_h^{+(n)} + b\left(x_h^{+(n)}+x_h^{0(n)}\right)\left(x_h^{+(n)}+x_c^{+(n)}\right) \right] }{ 1 + b }
\end{equation}
from which the remaining number fractions at time step $(n+1)$ are derived. Because our solution is implicit, the dimensionless
timestep $b$ can exceed unity (as it does for our runs
at low spatial resolution), and the system will still relax to its
correct equilibrium. This chemical equilibrium is discussed further in \S\ref{sec:density}.

Note that in contrast to H08 and E10, our calculations account for the
reverse reaction $H_h^0 + H_c^+ \rightarrow H_h^+ + H_c^0$.
Accounting for the reverse reaction helps us to avoid overestimating
the amount of hot neutral hydrogen. Our calculations of $n_h^0$ are
still overestimated, however, because we neglect thermal
equilibration, i.e., cooling of hot hydrogen by collisions with cold
hydrogen. In \S\ref{sec:future} we estimate the error incurred to be
on the order of unity.

\revised{
Our calculations of the neutral fraction in the mixing layer
do not explicitly account for photoionizations by Lyman
continuum photons, radiative recombinations, or advection of neutral hydrogen from the planetary wind --- but these effects are already included by \citet{Murray-Clay2009} whose planetary wind parameters we use; see \S\ref{sec:hydro}.
}





\subsection{Lyman-$\alpha$ Absorption}\label{sec:lyman}

The transmission spectrum in the Lyman-$\alpha$ line is 
post-processed, i.e., calculated after \texttt{HERACLES} has finished running.
Both hot and cold neutral hydrogen ($n^0_h$ and $n^0_c$) contribute
to the Lyman-$\alpha$ optical depth. It is assumed that the hot
and cold neutral hydrogen do not thermally equilibrate
(see \S\ref{sec:future} where we question this assumption). Thus in
computing the opacity due to hot hydrogen, we adopt a kinetic
temperature of $T_\ast = 10^6$ K, and in computing the opacity due
to cold hydrogen we take $T_p = 7000$ K. In each grid cell, the wavelength at line center is Doppler shifted according to the
horizontal component of the bulk velocity (the observer is to the far right
of the simulation box). Voigt line profiles are
used with a damping constant (Einstein A coefficient)
equal to $\Gamma = 6.365 \times 10^8$ s$^{-1}$ \citep[e.g.,][]{Verhamme2006}.

For each wavelength $\lambda$, the line-of-sight optical depth
$\tau_\lambda (y)$ is evaluated along each horizontal row of cells
pointing to the star (lying between the white dashed lines in Figures
\ref{r0050} and \ref{r3200}). The total absorption is then computed as
\begin{equation} \label{eqn:absorption}
A(\lambda) = \langle 1 - \exp (-\tau_\lambda) \rangle
\end{equation}
where $\langle \rangle$ denotes a 1-dimensional spatial average over $y$.
Of course the star actually presents a
circular disc, but because the simulation is only 2D, our simple 1D
average seems fair.  The absorption profile $A(\lambda)$ can be
computed for every snapshot (timestep) of the simulation.

\subsection{Differences Between This Work and E10/H08}\label{sec:difference}
The main difference between our methods and
those of E10/H08 is that we numerically solve the equations of
hydrodynamics in a 2D geometry, whereas E10/H08 simulate collisions of
hydrogen ``meta-particles'' in a more kinetic, 3D treatment.
Neither we nor they compute magnetic forces explicitly.

E10 include forces arising from the orbit of the planet about the
star, including the Coriolis force, the centrifugal force, and stellar
tidal gravity. We do not. Our focus is on resolving mixing and charge
exchange in the interface between the two winds. To that end, we solve
for both the forward and reverse reactions of charge exchange
(equations \ref{ce}--\ref{charge_exchange}), whereas E10/H08 
solve only for the forward reaction. Our equations permit a chemical
equilibrium to be established in the mixing layer; see
\S\ref{sec:density}. Furthermore, the structure and geometry of the
interaction region between the two winds are direct outcomes of our
simulations, whereas the shape of the interface layer is imposed as a fixed
``obstacle'' in the simulations of E10.

Other differences include our treatments of the planetary and stellar
winds. We account for both the neutral and ionized components of the
planetary wind; E10 assume the planetary outflow is purely neutral.
We draw our parameters of the stellar wind from those of the slow
equatorial Solar wind, which blows at $\sim$130 km/s at a
stellocentric distance of $r = 5R_\odot$ (Sheeley et al.~1997;
\citealt{quemerais2007}).  E10 take the stellar wind to blow at 450
km/s, while H08 take the stellar wind to blow at 50 km/s. Neither work
accounts for how the stellar wind decelerates due to its interaction
with the planetary wind, whereas in our simulations the deceleration
zones are well-resolved.

We will review again our simulation methods, and assess the severity
of our approximations, in \S\ref{sec:future}.

\section{RESULTS}\label{sec:results}

Results for Lyman-$\alpha$ absorption by the mixing layer,
including numerical convergence tests and a direct comparison
with observations, are given in \S\ref{resolution}.
A parameter study is described in \S\ref{sec:mixing}.

\subsection{Absorption vs.~Spatial Resolution and Time}\label{resolution}

\begin{figure*}
\centering
\includegraphics[width=0.32\linewidth]{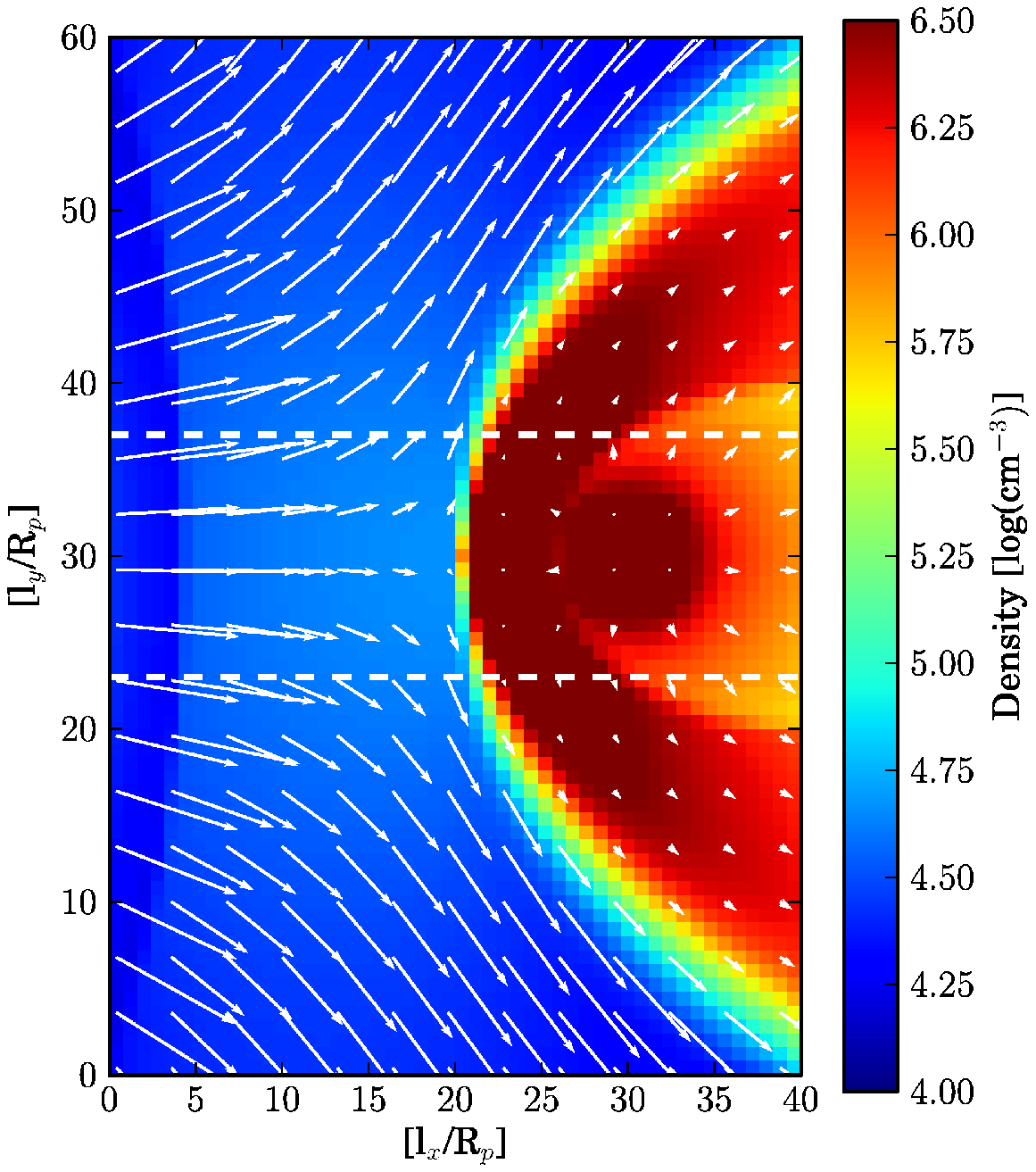}
\includegraphics[width=0.32\linewidth]{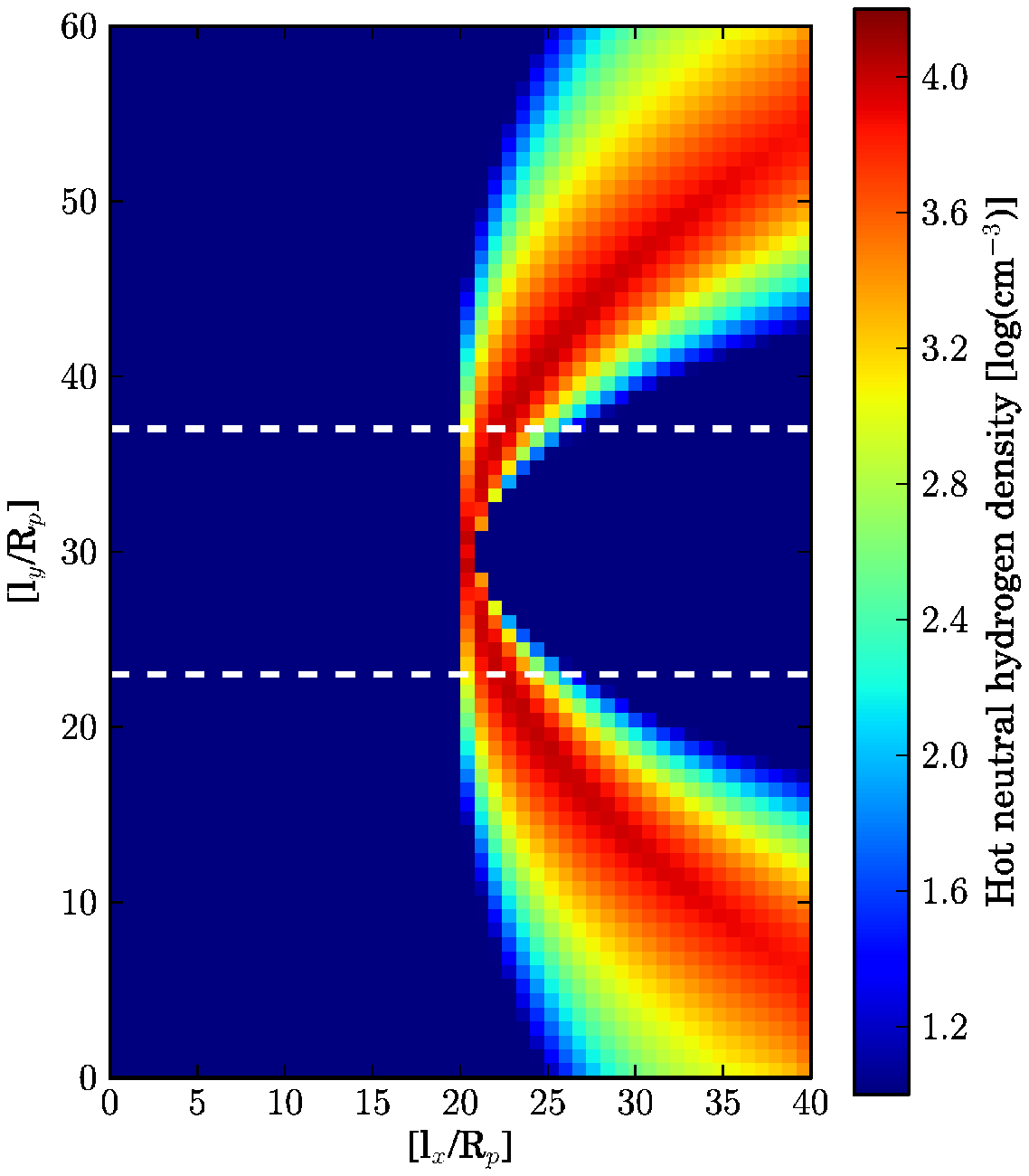}
\includegraphics[width=0.32\linewidth]{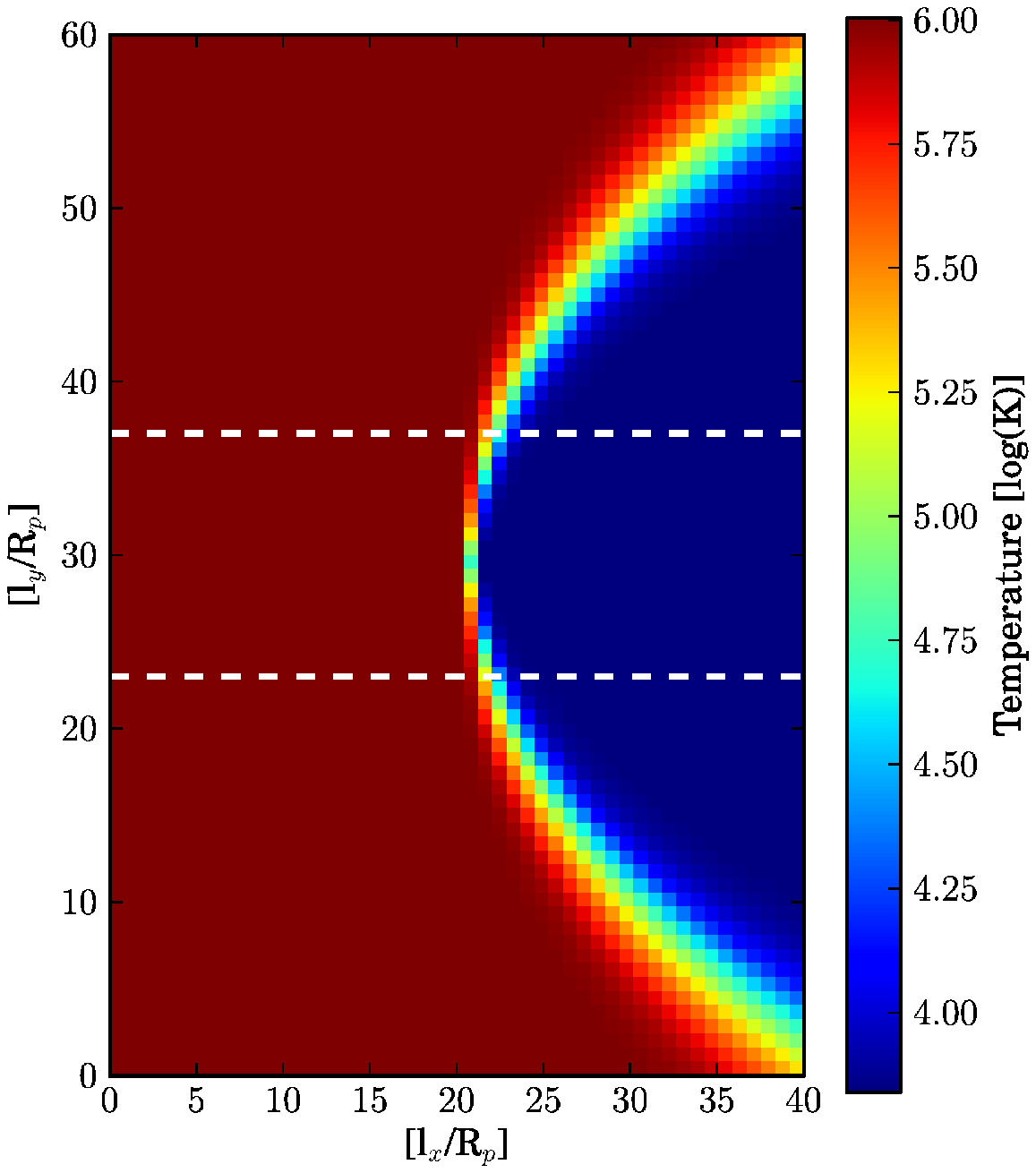}
\caption{Snapshots of total density and velocity (left panel), density of hot neutral hydrogen ($n_h^0$, middle panel), and temperature (right panel)
of the 50x75 simulation, with parameters listed in
  Table \ref{parameters}. Snapshots are taken at $t = 2\times 10^6$ s.
The temperature map shown in the right panel is computed by
\texttt{HERACLES} and used only to compute the hydrodynamic evolution;
it is not used to compute the charge exchange reactions
or the Lyman-$\alpha$ spectrum (see \S\ref{sec:ce}--\ref{sec:lyman}).
The two dashed white lines represent sightlines to the stellar limbs.}
\label{r0050}
\end{figure*}

\begin{figure*}
\centering
\includegraphics[width=0.32\linewidth]{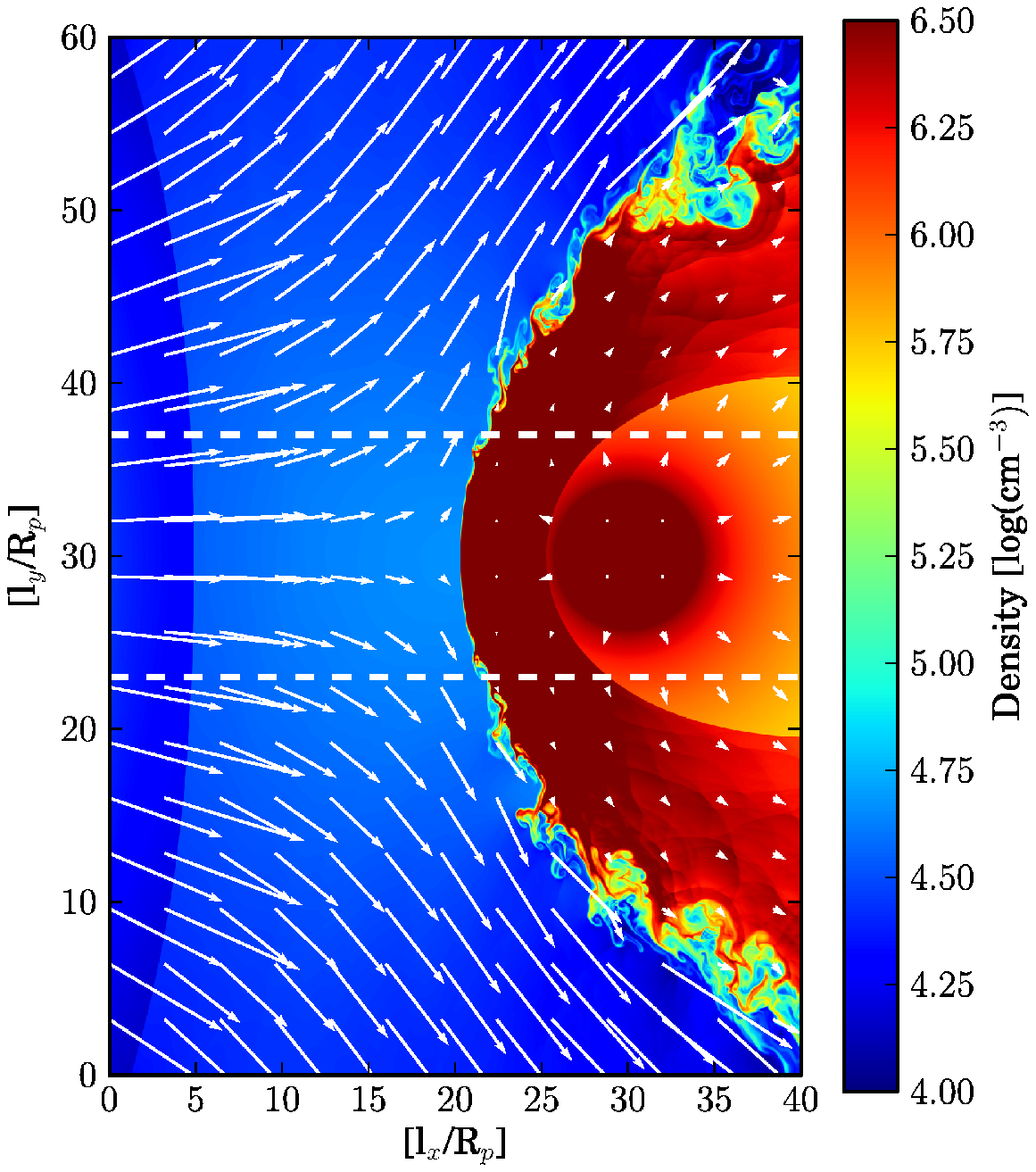}
\includegraphics[width=0.32\linewidth]{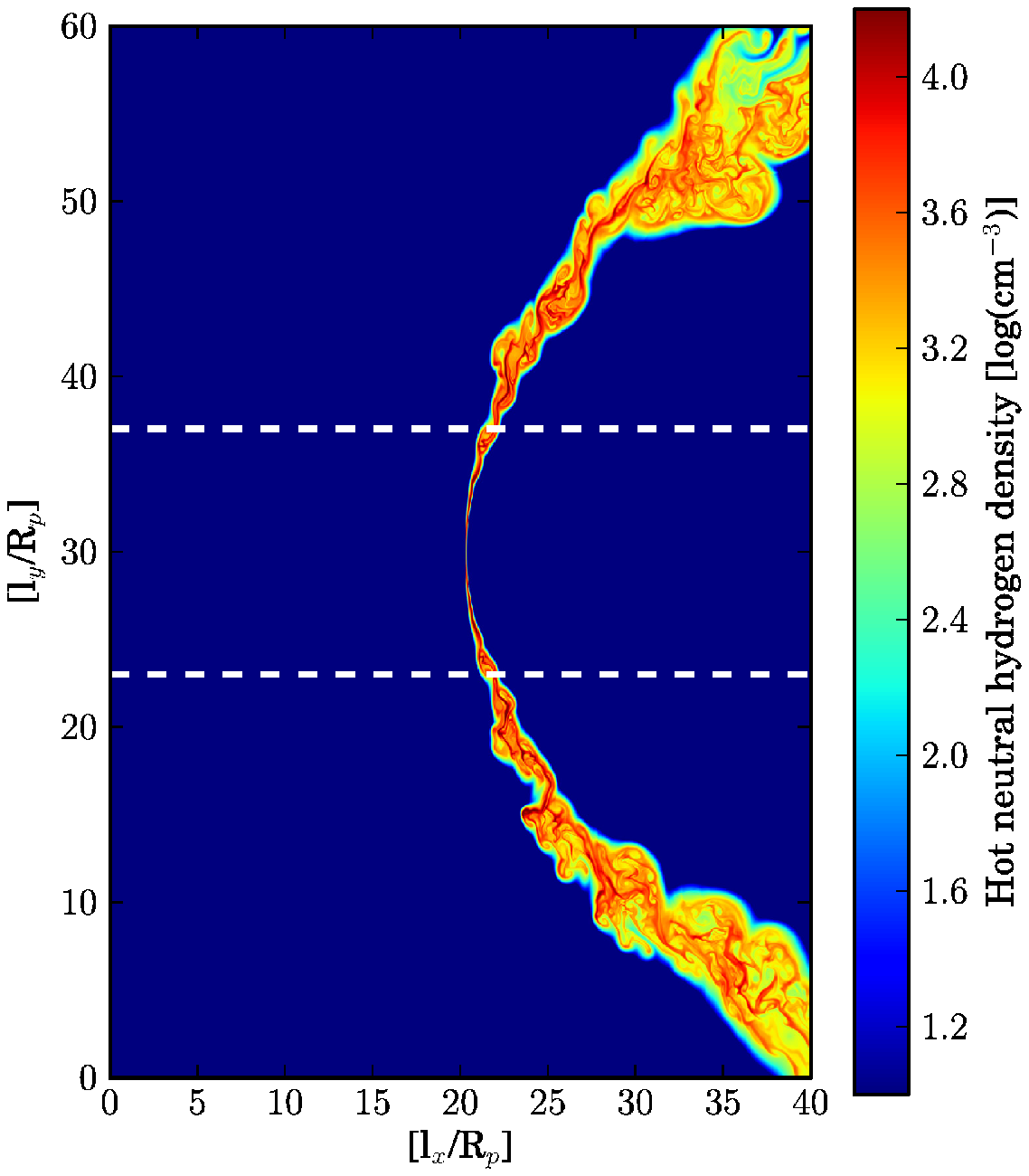}
\includegraphics[width=0.32\linewidth]{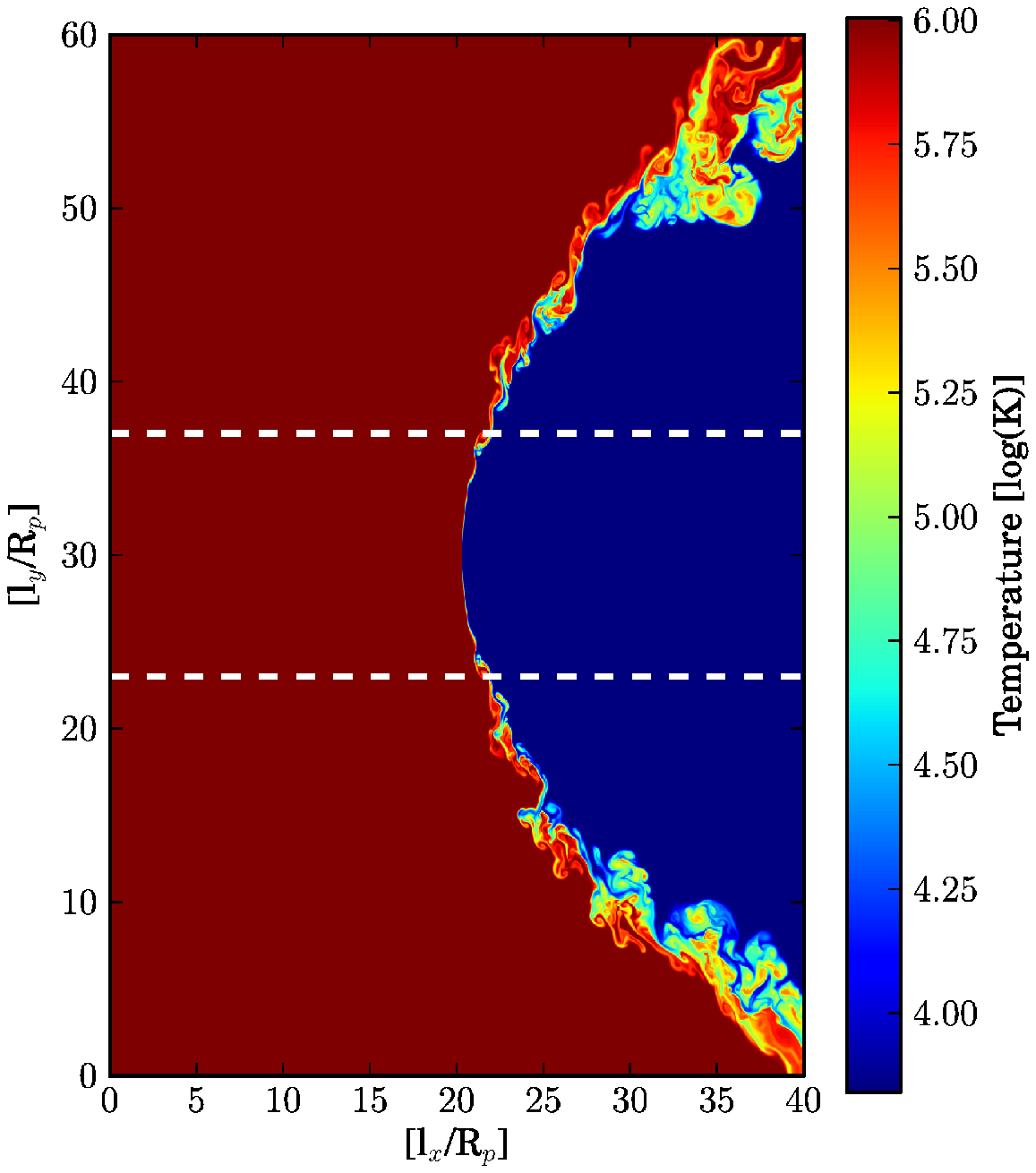}
\caption{Same as Figure \ref{r0050} and for the same simulation parameters but at a grid resolution of 3200x4800. Kelvin-Helmholtz rolls at the contact discontinuity first appear at this resolution. An even higher resolution of 6400x9600 yields the same star-averaged absorption; see Figure \ref{absorption_vs_time}.
}
\label{r3200}
\end{figure*}

In Figures \ref{r0050} and \ref{r3200}, we present results at our
lowest ($50 \times 75$) and near-highest ($3200 \times 4800$) spatial
resolutions, respectively. The simulations agree on the basic
properties of the flow. The planetary wind is launched from the red
circle and encounters a bow shock, visible in the left panels as a
curved boundary separating orange (unshocked planetary wind) from red
(shocked planetary wind).  The radius of curvature of the planetary
wind shock is roughly $\sim$$6R_p$. Outside, the red region of
thickness $\sim$$5R_p$ contains shocked planetary wind.

The stellar wind encounters a weak shock---visible as a near-vertical line
separating dark blue from lighter blue in the left-hand panels
of Figures \ref{r0050} and \ref{r3200}---at a distance of
$\sim$$5R_p$ from the left edge of the box. The shocked stellar
wind is diverted around the planet by the pressure
at the stagnation point where the two winds collide head on.

We observe that both winds accelerate somewhat before they encounter shocks.
For our standard model, 
\revised{
 the Mach numbers are
}
$M_\ast \lesssim 1.3$ and $M_p \lesssim 1.5$ 
(for the parameter study simulations of \S\ref{sec:mixing}, $M_p$
can grow up to 2--3). Density enhancements are thus modest---less than
a factor of 2.


The contact discontinuity between the stellar and planetary winds
separates light blue from dark red in the left panels. It is laminar at low
resolution but breaks up into turbulent Kelvin-Helmholtz rolls at high
resolution (cf. Stone \& Proga 2009 whose spatial
resolution was probably too low to detect the Kelvin-Helmholtz
instability). The middle panels plot the density of hot neutral
hydrogen produced by charge exchange in the mixing layer.  The
``head'' of the mixing layer, located near the stagnation point, spans
only one or two grid cells in the low resolution simulation. The high
resolution simulation resolves much better the head of the mixing
layer. Zoomed-in snapshots of the head will be presented in
\S\ref{sec:mixing}.



\begin{figure*}
\centering
\includegraphics[width=\linewidth]{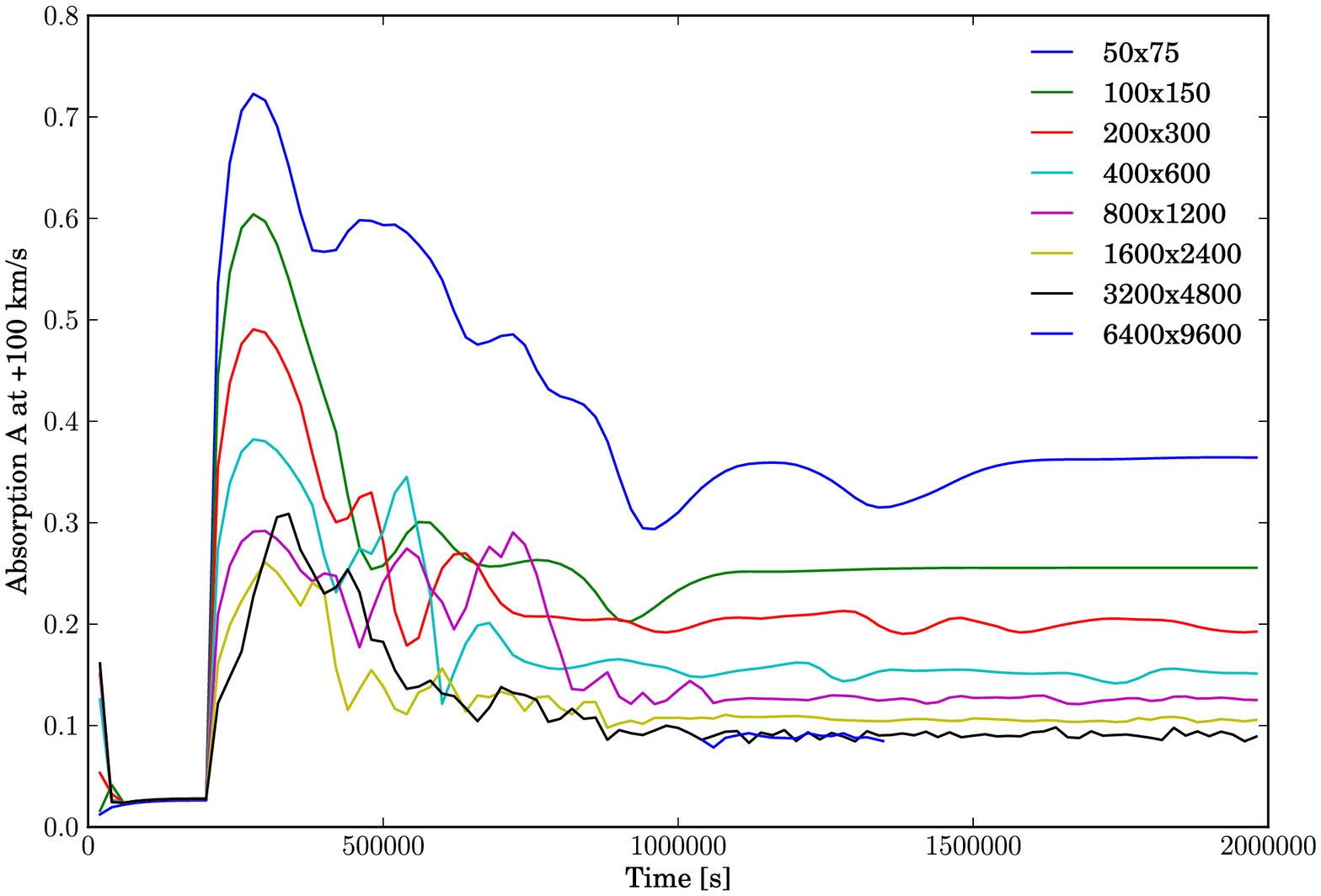}
\caption{Lyman-$\alpha$ absorption $A$ (equation \ref{eqn:absorption}), evaluated at a Doppler-shift velocity of +100 km/s from line center, versus time and spatial resolution. The absorption converges in time for all simulations, but only for grid resolutions of 3200x4800 or greater does a unique value for the absorption emerge. The 3200x4800 simulation is also the lowest resolution run to resolve Kelvin-Helmholtz billows; see Figure \ref{r3200}.}
\label{absorption_vs_time}
\end{figure*}

In Figure \ref{absorption_vs_time}, the star-averaged absorption $A$
at an equivalent Doppler velocity of +100 km/s (redshifted away from
the observer) is plotted against time for a range of spatial
resolutions. From $t = 0$ to $2 \times 10^5$ s, the planetary wind
fills the simulation domain; the absorption quickly settles down to a
value of $\sim$2\%. At these early times, only cold ($T_p = 7000$ K)
neutral hydrogen from the planet is available to absorb in
Ly-$\alpha$, and it is clearly insufficient to explain the absorption
observed with {\it HST}.

Starting at $t = 2 \times 10^5$ s, the stellar wind is injected into
the box. The absorption attains a first peak when the planetary and
stellar winds reach a rough momentum balance and a mixing layer
containing hot ($T_\ast = 10^6$ K) neutral hydrogen is established.
The height of the first peak decreases with each factor of 2
improvement in grid resolution until a resolution of 3200x4800 is
reached.  
Encouragingly, all of the absorption values calculated in the various
simulations converge at late times.


The 3200x4800 run is the best behaved, with the absorption
holding steady at $A \approx 9$\% for 10$^6$ s. Compared to all other
simulations at lower resolution, the 3200x4800 run is the only one in
which Kelvin-Helmholtz rolls appear (more on the Kelvin-Helmholtz
instability in \S\ref{sec:thickness}).

We further tested the convergence
of the 3200x4800 run by performing an even higher resolution simulation with
6400x9600 grid cells. Because of the expense of such a simulation, the
initial conditions of the 6400x9600 run were taken from the 3200x4800
run at $t = 10^6$ s, and integrated forward for only $3\times 10^5$ s
(approximately 9 box crossing times for the stellar wind in the
horizontal direction).  The absorption values versus time for the
6400x9600 run are overlaid in Figure \ref{absorption_vs_time} and are
practically indistinguishable from those of the 3200x4800 run.  Having
thus satisfied ourselves that the 3200x4800 run yields numerically
convergent results, we will utilize this grid resolution ($0.0125 R_p$
per grid cell length) for further experiments to understand the
dependence of the absorption on input parameters, as described in
\S\ref{sec:mixing}.

\begin{figure}
\centering
\includegraphics[width=\linewidth]{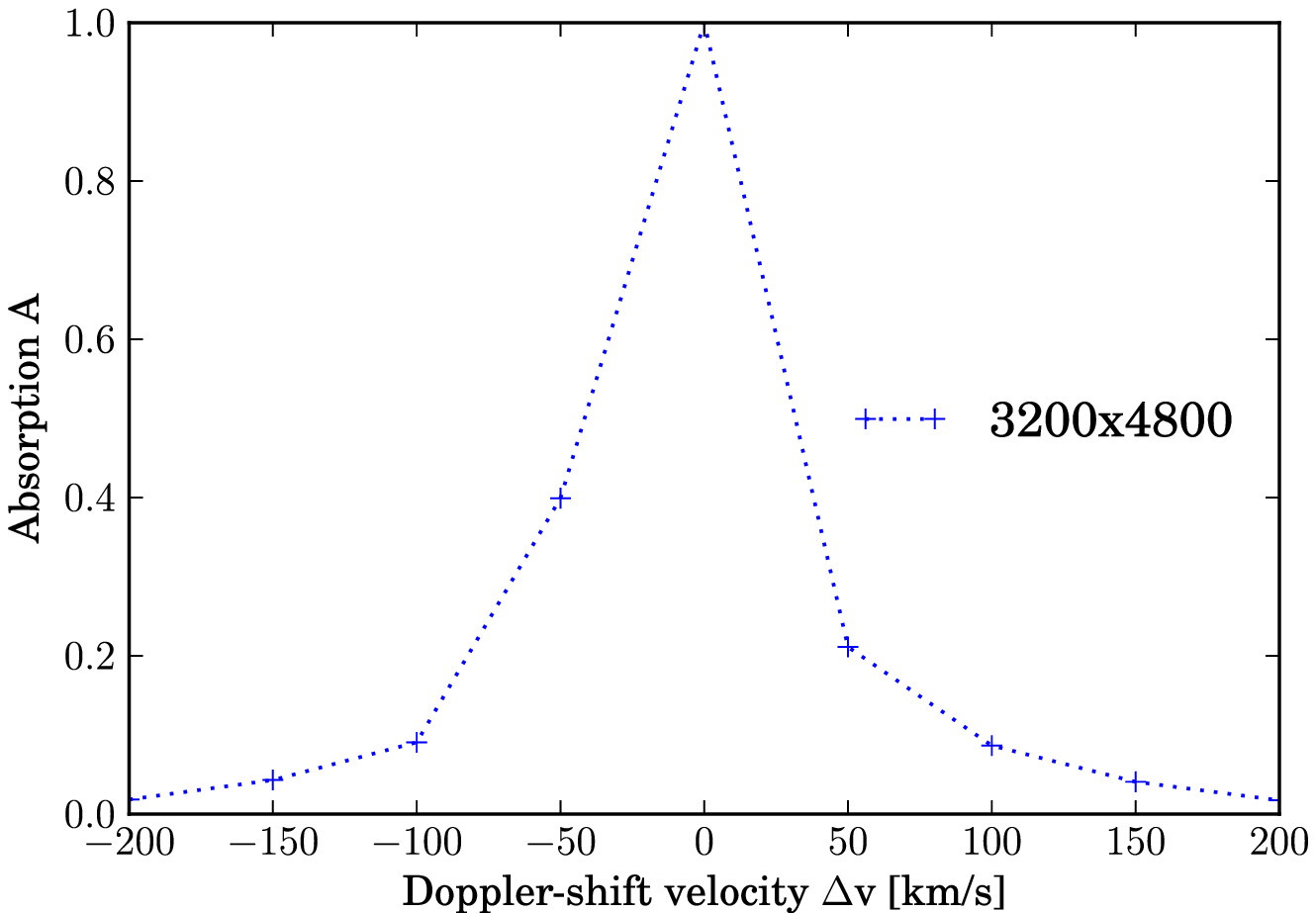}
\caption{Lyman-$\alpha$ absorption $A$ versus Doppler-shift velocity
  $\Delta v$ from line center, evaluated for our standard 3200x4800
  simulation at $t = 2\times 10^6$ s. Absorption at -50 km/s is
  stronger than at +50 km/s because of the bulk motion of
  charge-exchanged neutral hydrogen streaming from the star toward the
  observer. The line wings at larger Doppler shifts are primarily
  thermally broadened at $T_\ast = 10^6$ K. The absorption $A \approx
  9$\% at $\Delta v = \pm 100$ km/s, in accord with
  {\it HST} observations; see Figure \ref{lya_spectrum}.}
\label{absorption_spectrum}
\end{figure}

Figure \ref{absorption_spectrum} plots the absorption spectrum for our
standard 3200x4800 simulation at $t = 2 \times 10^6$ s.  The
absorption $A$ is evaluated at wavelengths offset from the central
rest-frame wavelength of the Lyman-$\alpha$ transition by 9 Doppler-shift
velocities $\Delta v$.  Absorption at -50 km/s is stronger than at
+50 km/s, a consequence of neutral, charge-exchanged hydrogen from the
star accelerating from the stagnation point toward the observer.  At
larger velocities $|\Delta v| > 100$ km/s, the spectrum is more nearly
reflection-symmetric about $\Delta v = 0$, because the broadening is
purely thermal at $T_\ast = 10^6$ K.

\begin{figure*}
\centering
\includegraphics[width=\linewidth]{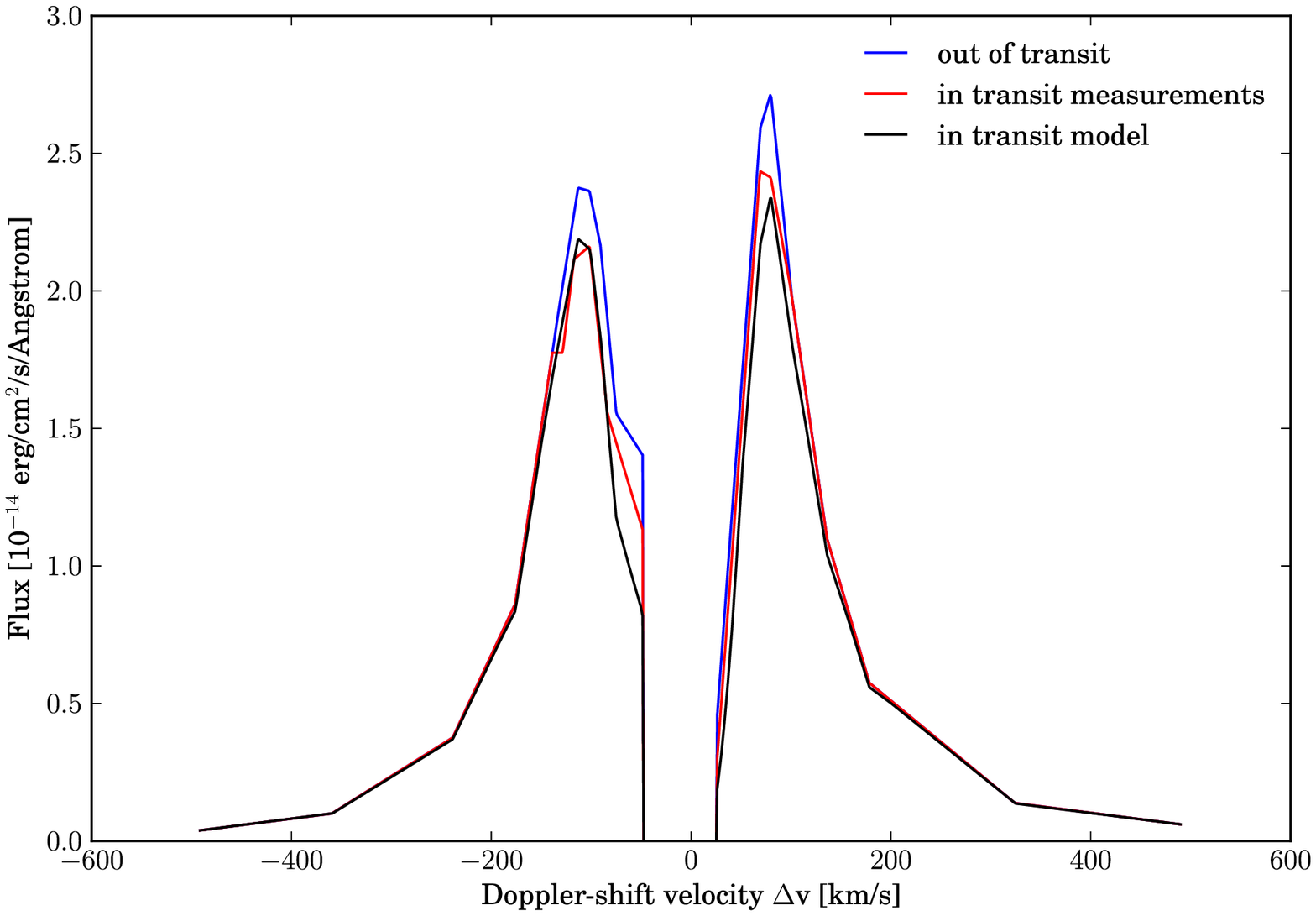}
\caption{Observed out-of-transit (highest blue curve) and observed
  in-transit (green curve) Lyman-$\alpha$ spectra, reproduced from
  Figure 2 of Vidal-Madjar et al.~(2003). In the line ``core'' from
  -42 to +32 km/s, where interstellar absorption is too strong to
  extract a planetary transit signal, the flux is set to zero. Our
  theoretical in-transit spectrum (red curve) is computed by
  multiplying the observed out-of-transit spectrum by $1-A$, where $A$
  is plotted in Figure \ref{absorption_spectrum}. The agreement
  between the theoretical and observed in-transit spectra is good,
  supporting the idea that charge exchange between the stellar and
  planetary winds correctly explains the observed absorption at Doppler-shift
  velocities around $\pm$100 km/s.}
\label{lya_spectrum}
\end{figure*}

Figure \ref{lya_spectrum} displays the same information as in
Figure \ref{absorption_spectrum} but in the full context
of the {\it Hubble Space Telescope} observations. The agreement
between the modeled and observed in-transit spectra is encouraging.

\begin{figure*}
\centering
\subfigure[Standard case]{\label{a}\includegraphics[trim = 2cm 0 0.cm 0,width=0.24\linewidth]{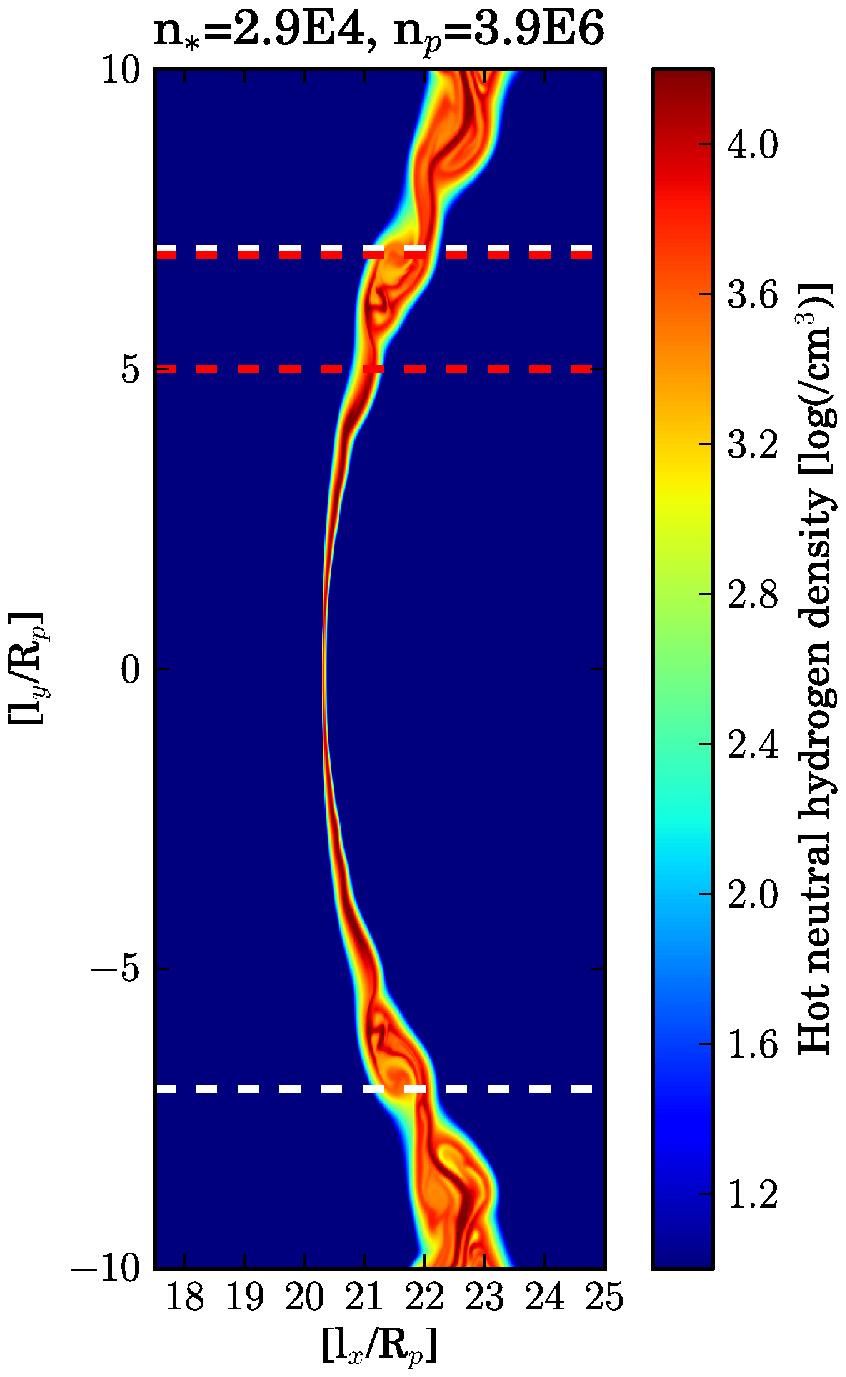}}
\subfigure[n$_p$ increased]{\label{b}\includegraphics[trim = 2cm 0 0.cm 0,width=0.24\linewidth]{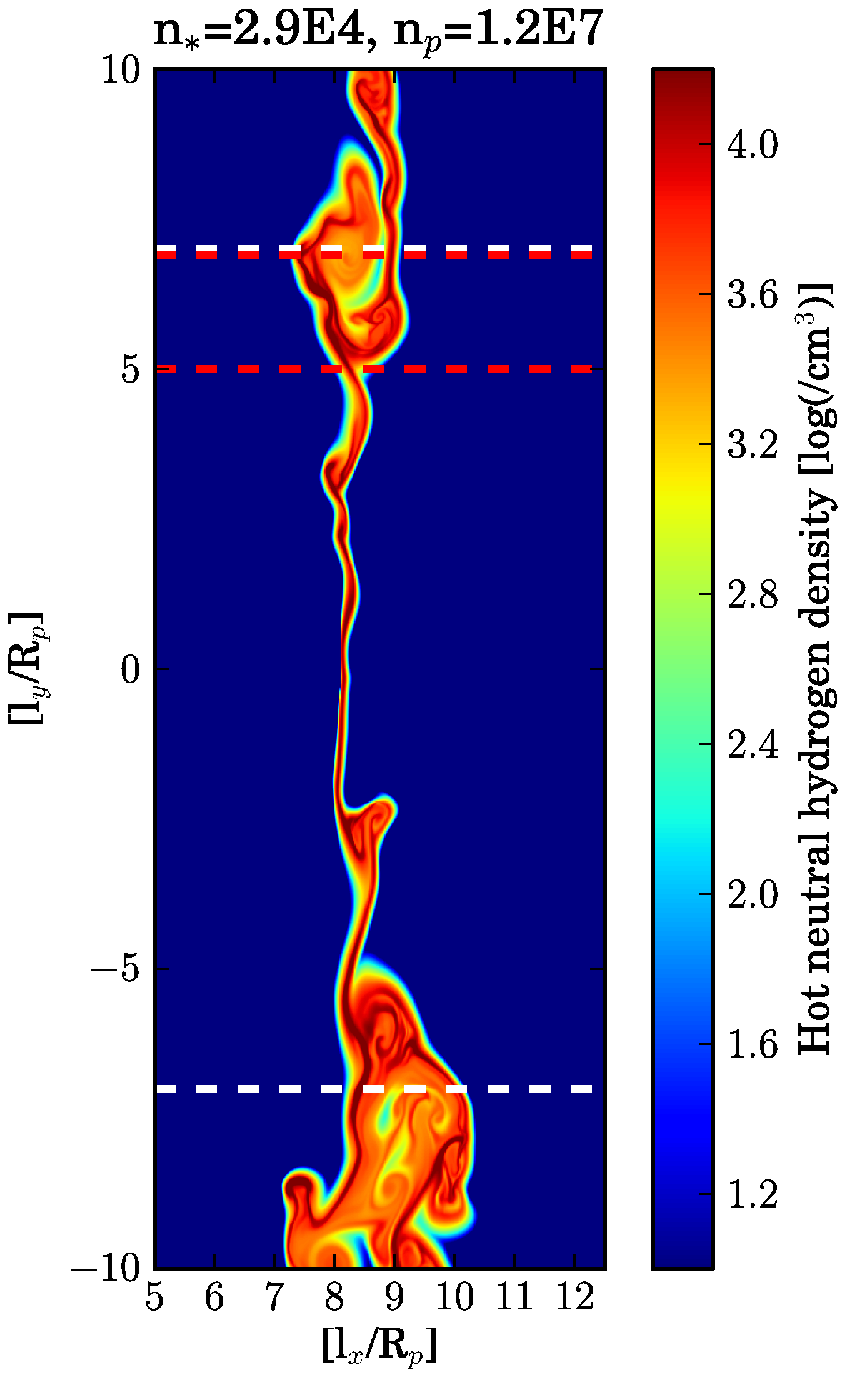}}
\subfigure[n$_\ast$ decreased]{\label{c}\includegraphics[trim = 2cm 0 0.cm 0,width=0.24\linewidth]{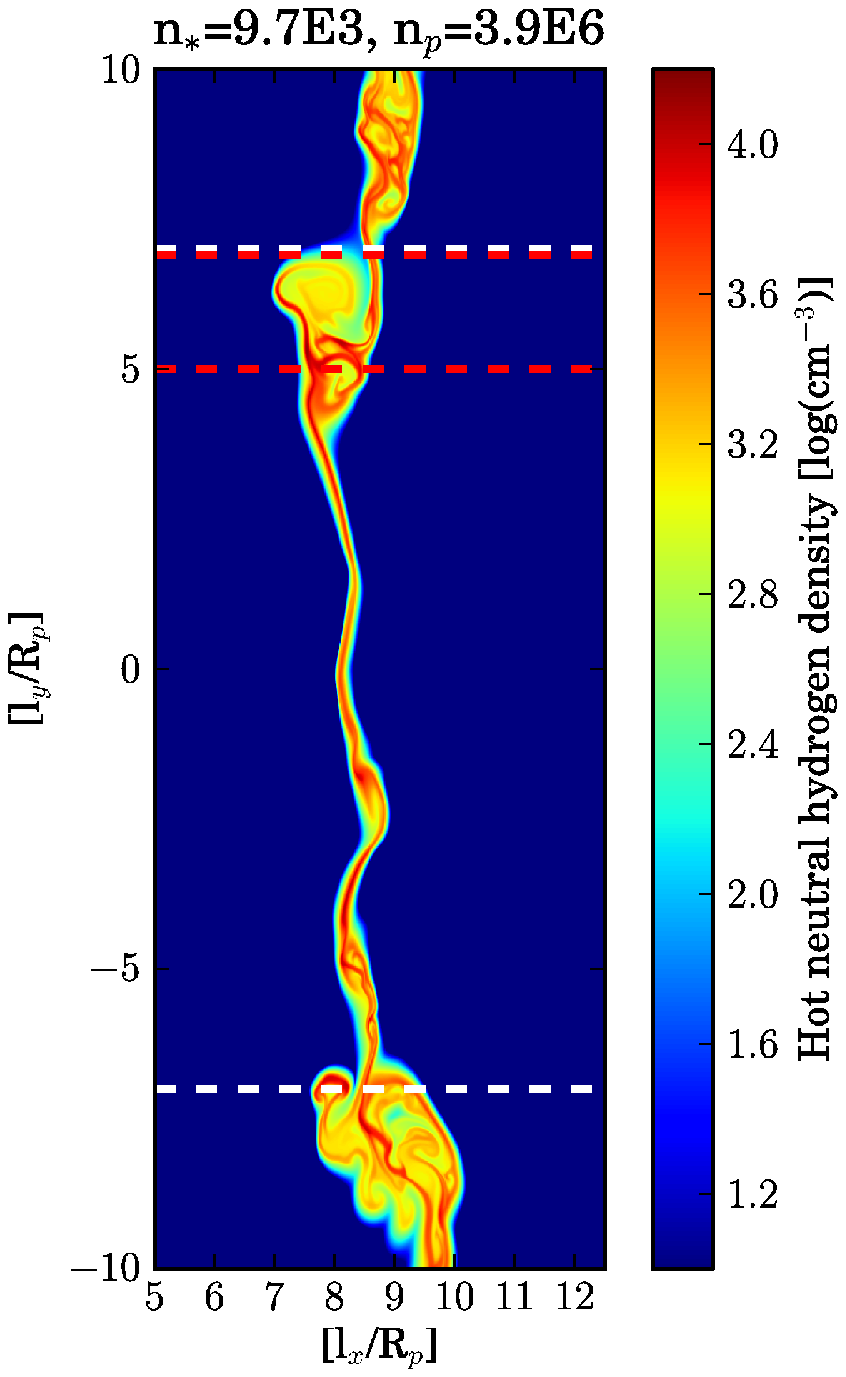}}
\subfigure[v$_p$,T$_p$ increased]{\label{d}\includegraphics[trim = 2cm 0 0.cm 0,width=0.24\linewidth]{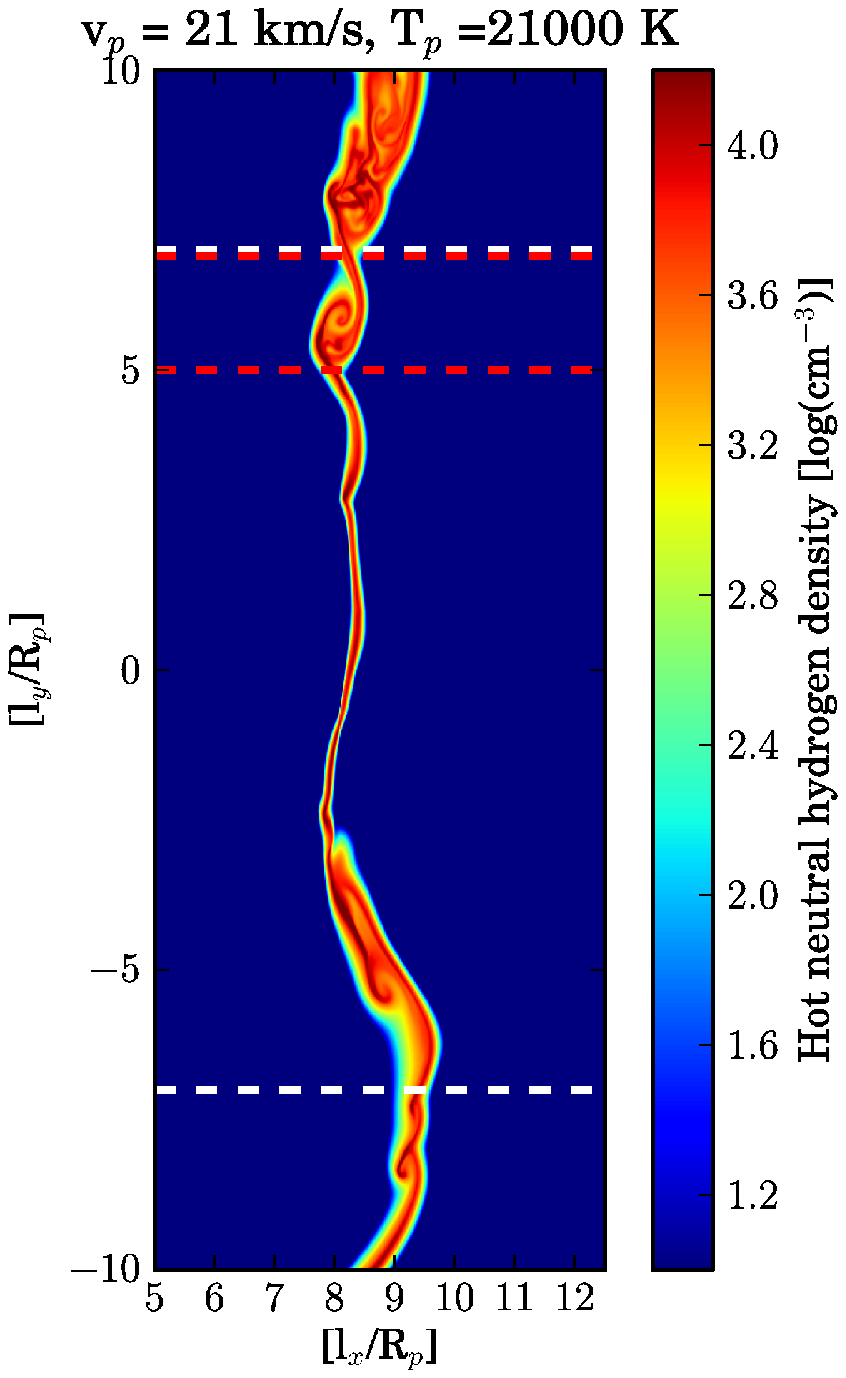}}
\caption{Zoomed-in snapshots of hot neutral hydrogen in the four simulations
used to study how the properties of the mixing layer depend on input parameters; see Table \ref{param_ml}. Snapshots are taken near $t = 2 \times 10^6$ for the standard model,
and near $t = 1\times 10^6$ s for the others. The bottom white dashed line is the line of sight to the lower stellar limb. The upper two red dashed lines bracket the ``sampling interval'' over which the hot neutral hydrogen density is vertically averaged to produce the density profiles shown in Figure \ref{cn_profiles} (the uppermost red dashed line is also the sightline to the upper stellar limb). In cases (b), (c), and (d), the mixing layer is located farther from the planet (centered at $l_x = 30R_p$) than is the case in (a). In cases (b) and (c), the horizontal thickness of the mixing layer is greater than in cases (a) and (d). And in case (c), the density of hot neutral hydrogen is lowest. See \S\ref{sec:mixing} for explanations.
}
\label{mixing_layer}
\end{figure*}

%
%

\subsection{Scaling Relations for Absorption in the Mixing Layer} \label{sec:mixing}

To understand how absorption in the mixing layer depends on input
parameters, we performed 3 additional simulations varying the launch
properties $n_p$, $n_\ast$, $v_p$ and $T_p$. The altered parameters
are listed in Table \ref{param_ml}. For all 3 simulations, the box size
was maintained at $(L_x, L_y) = (40R_p, 60 R_p)$ and the grid resolution
was $(N_x, N_y) = (3200, 4800)$.

Note that our parameter study is not exhaustive. For example,
none of the simulations listed in Table \ref{param_ml} varies
the Mach number at launch of either the planetary or stellar wind.
Actually we have performed simulations varying the Mach number
of the stellar wind. These behave as we would expect---in particular,
increasing $M_\ast$ increases the amount of absorption because
of the increased compression in the stellar shock. Nevertheless
we elect not to include these extra simulations
in our parameter study below. Magnetic fields, neglected
by our simulations but certainly present in the stellar wind
if not also the planetary wind, would stiffen the gas and prevent
the kind of compression that we see when we raise $M_\ast$.

In the following subsections, we explain our numerical results on the
properties of the mixing layer with order-of-magnitude scaling
relations.  The mixing layer's location is analyzed in
\S\ref{sec:location}; its thickness in \S\ref{sec:thickness}; the
densities of its constituent species in \S\ref{sec:density}; and the
column density and absorptivity of hot neutral hydrogen in
\S\ref{sec:column}.

\begin{table}
\caption{\label{param_ml} Launch parameters for the 3 additional $40R_p \times 60 R_p$ simulations at $3200 \times 4800$ resolution. The stellar parameters $v_*$ and $T_*$ are kept at their nominal values from Table \ref{parameters}. Note that none of the Mach numbers change.}
\centering
\begin{tabular}{l|l||l|l|l}

     \hline
\hline
&Nominal & n$_p \uparrow$ & n$_* \downarrow $ & v$_p$,T$_p \uparrow$ \\
\hline
n$_*$  & 2.9E4/cm$^3$ & 2.9E4/cm$^3$ & 9.7E3/cm$^3$ & 2.9E4/cm$^3$ \\
n$_p$  & 3.9E6/cm$^3$ & 1.2E7/cm$^3$ & 3.9E7/cm$^3$ & 3.9E6/cm$^3$ \\

v$_p$  & 12 km/s & 12 km/s & 12 km/s & $\sqrt{3}\times$12 km/s\\
T$_p$  & 7000 K &  7000 K &  7000 K &  21000 K \\
\hline
$\mathcal{R}$&$\mathcal{R}_0$ = 0.11 & $3\times \mathcal{R}_0$ & $3 \times \mathcal{R}_0$ & $3 \times \mathcal{R}_0$ \\
\end{tabular}
\end{table}

\subsubsection{Location of the mixing layer}\label{sec:location}
Along the line joining the planet to the star, the mixing layer---equivalently,
the contact discontinuity---is located approximately where the two winds reach
pressure balance:
\begin{equation}\label{eqn:pressure}
\rho_\ast (v_\ast^2 + c_\ast^2) = \rho_p (v_p^2 + c_p^2) \,.
\end{equation}
In equation (\ref{eqn:pressure}), quantities are evaluated near the
mixing layer, not at launch. Note further that in equation
(\ref{eqn:pressure}) and in equations to follow, we ignore the
distinction between shocked and unshocked gas, as wind Mach numbers
are near unity.  Idealizing each wind velocity as constant, we
substitute $\rho_p = \dot{M}_p/(2\pi v_p d_p)$ and $\rho_\ast =
\dot{M}_\ast/(2\pi v_\ast d_\ast)$ into equation (\ref{eqn:pressure}),
as appropriate for the 2D circular winds in our simulations.  Here
$d_p$ measures distance from the planet, and $d_\ast$ measures
distance from the star. Then the distance from the planet to the
mixing layer---i.e., the approximate radius of curvature of the mixing
layer---is given by
\begin{equation} \label{location_ml}
d_p = d_\ast \mathcal{R}
\end{equation}
where
\begin{equation}
 \mathcal{R} \equiv \frac{\dot{M}_p (v_p^2 + c_p^2) / v_p}{\dot{M}_\ast (v_\ast^2 + c_\ast^2) / v_\ast} \,.
\end{equation}
For our standard model, $\mathcal{R} = \mathcal{R}_0 \approx 0.11$. 
\revised{
Note that for 3D
spherical winds, $d_p = d_\ast \sqrt{\mathcal{R}}$
\citep{Stevens:1992gf}, but this relation is not relevant for our 2D
Cartesian simulations --- we will use (\ref{location_ml}) instead.
}



The parameters in Table \ref{param_ml} were chosen to increase
$\mathcal{R}$ by a factor of 3 compared to its value in our fiducial
model. By equation (\ref{location_ml}), when $\mathcal{R} = 3
\mathcal{R}_0$, the mixing layer should be displaced $3\times$ farther
away from the planet compared to its location in our standard model,
assuming the star is far enough away that $d_\ast$ is essentially
fixed at the star-planet separation. Figure \ref{mixing_layer}
displays zoomed-in snapshots of the mixing layers for all simulations
in Table \ref{param_ml}. Looking at the $l_x$-positions of the mixing
layers, and recalling that the planet sits at $l_x = 30R_p$, we find that
the layer is displaced $(30-8)/(30-21) \approx 2.4 \times$ farther
away in the three new simulations as compared to the standard model.
We consider this close enough to our expected factor of 3,
given our neglect of the rather
thick layers of shocked gas surrounding the mixing layer.

\begin{figure}
\centering
\includegraphics[width=\linewidth]{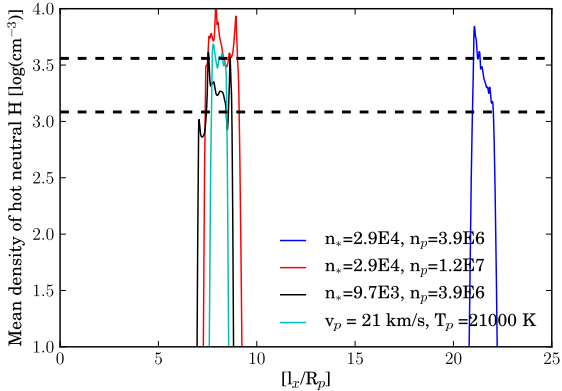}
\caption{Density profiles of charge-exchanged hot neutral hydrogen in
  the mixing layer, averaged vertically (between the dashed red lines
  in Figure \ref{mixing_layer}) and plotted against horizontal
  position. The planet is located to the right at $l_x = 30R_p$. The
  mixing layers of the three non-standard simulations are all
  displaced farther from the planet than in the standard model, a
  consequence of increasing the ratio $\mathcal{R}$ of the momentum
  carried by the planetary wind to that of the stellar wind
  (\S\ref{sec:location}).  The thicknesses of the mixing layers as
  shown by the red and green curves are larger than those shown by the
  blue and cyan curves, a consequence of changing the growth rate for
  the Kelvin-Helmholtz instability (\S\ref{sec:thickness}).  The two
  dashed lines are predictions of equation (\ref{hot_neutral_density})
  based on considerations of chemical equilibrium; the blue, green,
  and cyan curves correctly intersect the upper dashed line, while the
  red curve correctly intersects the lower dashed line
  (\S\ref{sec:density}).}
\label{cn_profiles}
\end{figure}

Figure \ref{cn_profiles} shows density profiles for hot neutral hydrogen
in the mixing layer for the three simulations plus our standard model.
Densities are averaged over $l_y$
and plotted against $l_x$. The fact that the mixing layers in the simulations
having $\mathcal{R} = 3 \mathcal{R}_0$ align in position confirms that
$\mathcal{R}$ is the dimensionless parameter controlling the location
of the mixing layer.

\subsubsection{Thickness of the mixing layer} \label{sec:thickness}
Figure \ref{cn_profiles} also indicates that the thickness of the
mixing layer, $L_{\rm mix}$, varies when we change input parameters.
Empirically, we find that the variations are consistent with
the relation
\begin{equation} \label{mix_length}
L_{\rm mix} \sim 0.1R_p \left( \frac{n_p}{n_\ast} \right)^{0.5} 
\end{equation}
where, as before, the distinction between shocked and unshocked gas
densities is ignored.


We can rationalize (\ref{mix_length}) as follows.  The timescale for a
mode of wavelength $\lambda_{\rm KH}$ 
to grow exponentially by the linear Kelvin-Helmholtz instability (KHI)
is given by
\begin{equation} \label{linear_time}
t_{\rm KH} \sim \frac{\lambda_{\rm KH}}{v_\ast - v_p} \frac{\rho_p+\rho_\ast}{2\pi(\rho_p\rho_\ast)^{1/2}} \sim \frac{\lambda_{\rm KH}}{2\pi v_\ast} \sqrt{\frac{\rho_p}{\rho_\ast}} 
\end{equation}
(e.g., \citealt{chandra61}).
We assume that the thickness of the mixing layer saturates when a
certain mode first becomes nonlinear.  Near saturation, the velocity
perpendicular to the background shear flow becomes comparable to the
shear flow velocity: $v_\perp \sim v_\ast - v_p \sim v_\ast$.
Thus when the mode becomes nonlinear, the mixing layer has thickness
$L_{\rm mix} \sim v_\perp t_{\rm KH} \sim (\lambda_{\rm KH}/2\pi)
\sqrt{\rho_p/\rho_\ast}$. This result matches 
(\ref{mix_length}), if we assume the initial disturbance that develops
into the mixing layer has a characteristic lengthscale that is fixed
at $\lambda_{\rm KH} \sim R_p$. Our description of mode saturation
can only apply to locations not too far downstream from the
stagnation point; far away, the flows are too strongly perturbed to be
described by the linear growth timescale (\ref{linear_time}).

\subsubsection{Density of hot neutral hydrogen in the mixing layer}\label{sec:density}
The density of hot neutral hydrogen in the mixing layer is set by chemical
equilibrium. Suppose that within the layer, the total density
$n_{{\rm H,mix}}$
is approximately the average of the planetary wind density
and the stellar wind density:
\begin{equation} \label{half}
n_{{\rm H,mix}} \sim \frac{n_p + n_\ast}{2} \sim \frac{n_p}{2} \,.
\end{equation}
The densities in (\ref{half}) are those of shocked gas, but
as is the case for all of \S\ref{sec:mixing}, we ignore for simplicity
the difference in density between pre-shock and post-shock gas (see \S\ref{sec:location}).
Because $n_p \gg n_\ast$, charge exchange hardly alters the ionization
state of the shocked---and still cold---planetary wind. That is, the
values of $x_c^0$ and $x_c^+$ do not change as the dense planetary wind
mixes with the dilute stellar wind. In particular,
the ratio $x_c^0/x_c^+$ is fixed at its initial
value of $(1-f_p^+)/f_p^+ = 1/4$.

The timescale for charge exchange is $(n_{{\rm H,mix}} \beta)^{-1}
\sim 10$ s, much shorter than the hours required for stellar-occulting gas to
travel from the stagnation point to regions off the projected
stellar limb.
Thus nearly all of the gas seen in transit is driven quickly into
chemical equilibrium, which from equation (\ref{charge_exchange}) demands that:
\begin{align}
\frac{x^0_h}{x^+_h} &= \frac{x^0_c}{x^+_c} \\
&= \frac{1-f_p^+}{f_p^+} = \frac{1}{4} \,. \label{fix_ratio}
\end{align}
In other words, in the mixing layer, the ionization fraction of stellar wind material quickly
slaves itself to the ionization fraction of planetary wind material.
Now all of the hot hydrogen (both neutral and ionized) in the mixing layer
originates from the stellar wind; from (\ref{half}), we have
\begin{equation} \label{all_hot}
\left( x_h^0 + x_h^+ \right) n_{\rm H,mix} \sim \frac{n_\ast}{2} \,.
\end{equation}
Combining (\ref{fix_ratio}) with (\ref{all_hot}) yields
\begin{equation} \label{hot_neutral_density}
n_h^0 = x_h^0 n_{\rm H,mix} \sim \frac{1}{2} \left( 1-f_p^+ \right) n_\ast \,.
\end{equation}
Equation (\ref{hot_neutral_density}) is approximately
confirmed by our numerical results
in Figure \ref{cn_profiles}; the horizontal dashed lines predicted
by (\ref{hot_neutral_density}) roughly match
the densities from our numerical simulations.

\subsubsection{Column density and absorptivity of hot neutral hydrogen}\label{sec:column}
Combining (\ref{mix_length}) with (\ref{hot_neutral_density}) gives the total
column density of hot neutral hydrogen:
\begin{equation} \label{eqn:column}
N_h^0 \sim n_h^0 L_{\rm mix} \sim 
0.05 \left( 1-f_p^+ \right) \left( n_p n_\ast \right)^{1/2} R_p \,.
\end{equation}
For our standard model, $N_h^0 \sim 3 \times 10^{13}$ cm$^{-2}$.

During planetary transit, the hot absorbing gas that covers
the face of the star is located near the stagnation point.
As such, the bulk line-of-sight velocity of transiting gas is much less
than its thermal velocity, which is of order 100 km/s.
Assuming that the gas is only
thermally broadened, and that the gas is optically thin at wavelengths
Doppler-shifted from line center by velocities $\Delta v$,
we construct an approximate, semi-empirical formula
for the absorption:
\begin{align}
A(\Delta v) \sim & \,\, N_h^0 \sigma_{\rm line-ctr} \exp[ -m_H(\Delta v)^2/2kT_\ast ] \\
  \sim & \,\, 0.1 \left( \frac{1-f_p^+}{0.2} \right) 
\left( \frac{n_p}{4\times 10^6 \,{\rm cm}^{-3}} \right)^{1/2}\left( \frac{n_\ast}{3\times 10^4 \,{\rm cm}^{-3}} \right)^{1/2}\nonumber \\
   & \times  \left( \frac{10^6 \, {\rm K}}{T_\ast} \right)^{1/2} \left( \frac{ \exp[ -m_H(\Delta v)^2/2kT_\ast ]}{0.5} \right) \label{eqn:master}
\end{align}
where $\sigma_{\rm line-ctr} = 6 \times 10^{-15} (10^6 \, {\rm
  K}/T_\ast)^{1/2} \, {\rm cm}^2$ is the line-center cross section for
the Lyman-$\alpha$ transition, $m_H$ is the mass of the hydrogen atom,
and $k$ is Boltzmann's constant.  
Strictly speaking, the quantities in equation (\ref{eqn:master})
should be evaluated in the vicinity of the contact
discontinuity, but we have instead normalized equation (\ref{eqn:master})
to the wind properties at launch (evaluated at $d_{\rm launch,p} = 4R_p$
and $r_{{\rm launch},\ast} = 5R_{\odot}$). We have verified in our simulations
that the launch properties
differ only by factors of order unity from the values
at the contact discontinuity, and so equation (\ref{eqn:master})
may be used to predict the absorption by inserting only the launch properties.
The exponential in equation (\ref{eqn:master}) is evaluated for nominal
parameters $\Delta v = 100$ km/s and $T_\ast = 10^6$ K.

\begin{figure}
\centering
\includegraphics[width=\linewidth]{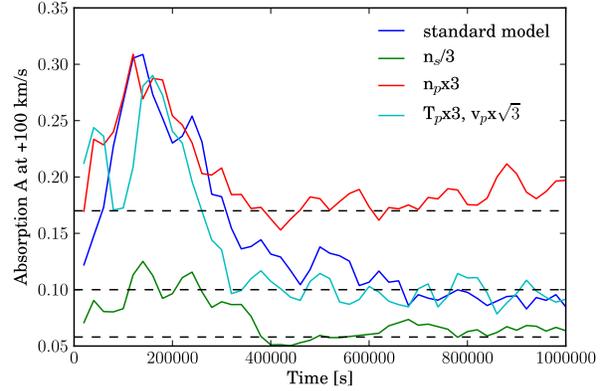}
\caption{Lyman-$\alpha$ absorption $A$ versus time, evaluated at a
  Doppler-shift velocity of +100 km/s from line center, for our
  standard model plus three additional models with different input
  parameters as indicated in the legend (see also Table
  \ref{param_ml}). The colored jagged lines are the results from our
  numerical simulations. The dashed black lines are the predictions
  from our physically motivated scaling relation (\ref{eqn:master});
  the simulations converge fairly well to the predicted values.}
\label{absorption_extra}
\end{figure}

As a further check, we show in Figure \ref{absorption_extra} the absorption
values $A$ to which the four simulations converge. They compare well
with the values predicted by (\ref{eqn:master}) using only the launch
properties.

Had we kept the dependence of the mixing layer properties on the
stellar wind Mach number $M_\ast$, equation
(\ref{hot_neutral_density}) would be modified such that $n_h^0 \propto
M_\ast^2 n_\ast$---where $n_\ast$ is the pre-shock (launch)
density---in accord with the usual Rankine-Hugoniot jump condition
that states that the density increases by the square of the Mach number
across a plane-parallel isothermal shock. And equations
(\ref{eqn:column})--(\ref{eqn:master}) would be modified such that $A
\propto N_h^0 \propto M_\ast$.  Indeed our numerical simulations (not
shown) confirm this linear dependence of $A$ on $M_\ast$.  We mention
this result only in passing because it is not likely to remain true
once we account for the real-life magnetization of the stellar
wind. Magnetic fields stiffen gas and reduce the dependence of $A$ on $M_\ast$.

\section{SUMMARY AND DISCUSSION}\label{sec:summary}
Using a 2D numerical hydrodynamics code, we simulated the collisional
interaction between two winds, one emanating from a hot Jupiter and
the other from its host star. The winds were assumed for simplicity to
be unmagnetized. Properties of the stellar wind were drawn directly
from observations of the equatorial slow Solar wind (Sheeley et
al.~1997; Qu\'emerais et al.~2007; Lemaire 2011), while those of the
planetary wind were taken from hydrodynamic models of outflows powered
by photoionization heating (Garc\'ia-Mu\~noz 2007; Murray-Clay et
al.~2009). For our standard parameters, the mass loss rate of the star
is $\dot{M}_\ast = 2 \times 10^{-14} M_{\odot}$/yr $ = 10^{12}$ g/s and
the mass loss rate of the planet is $\dot{M}_p = 1.6 \times 10^{11}$
g/s $= 2.7 \times 10^{-3} M_{\rm J}$/Gyr. At the relevant distances,
each wind is marginally supersonic---the stellar wind blows at $\sim$130--170 km/s
(sonic Mach number $M_\ast \lesssim 1.3$) and the planetary wind
blows at $\sim$12--15 km/s (Mach number $M_p \lesssim 1.5$). Thus shock
compression is modest, even without additional stiffening of the gas
by magnetic fields.

A strong shear flow exists at the contact discontinuity between the
two winds. At sufficiently high spatial resolution, we observed the
interfacial flow to be disrupted by the Kelvin-Helmholtz
instability. The Kelvin-Helmholtz rolls mix cold, partially neutral
planetary gas with hot, completely ionized stellar gas. Charge
exchange in the mixing layer produces observable amounts of hot
($10^6$ K) neutral hydrogen. Upon impacting the planetary wind, the
hot stellar wind acquires, within tens of seconds, a neutral component
whose fractional density equals the neutral fraction of the planetary
wind (about $1-f_p^+ = 20$\%). Seen transiting against the star, hot
neutral hydrogen in the mixing layer absorbs $\sim$10\% of the light
in the thermally broadened wings of the stellar Lyman-$\alpha$
emission line, at Doppler shifts of $\sim$100 km/s from line
center. Just such a transit signal has been observed with the {\it
  Hubble Space Telescope} (Vidal-Madjar et al.~2003).
\revised{
The $\pm$100 km/s velocities reflect the characteristic
velocity dispersions of protons in the stellar wind ---
as inferred from in-situ spacecraft observations of the Solar wind (e.g., Figure 3 of \citealt{Marsch06}).}

Our work supports the proposal by Holmstr\"om et al.~(2008) and
Ekenb\"ack et al.~(2010) that charge exchange between the stellar and
planetary winds is responsible for the Ly-$\alpha$ absorption observed
by {\it HST}. \revised{
This same conclusion is reached by \citet{LecavelierdesEtangs:2012jq}
in the specific case of HD 189733b.} 
Our ability to reproduce the observations corroborates
the first-principles calculations of hot Jupiter mass loss on which we
have relied (e.g., Yelle 2004; Garc\'ia-Mu\~noz 2007; Murray-Clay et
al.~2009, M09). Time variations in Ly-$\alpha$ absorption are expected both
from the variable stellar wind --- the Solar wind is notoriously
gusty --- and from the variable planetary wind, whose mass loss rate
tracks the time-variable ultraviolet and X-ray stellar luminosity.

\subsection{Neglected Effects and Directions for Future Research}\label{sec:future}

Although the general idea of photoionization-powered planetary
outflows exchanging charge with their host stellar winds seems
correct, details remain uncertain.
We list below some unresolved
issues, and review the effects that our simulations have neglected, in order
of decreasing concern.

\begin{enumerate}

\item {\it Thermal equilibration in the mixing layer.}  Our
  calculations overestimate the amount of hot neutral hydrogen
  produced by charge exchange because they neglect thermal
  equilibration.  A hot neutral hydrogen atom cools by colliding with
  cold gas, both ionized and neutral, from the planetary wind.
The concern is that hot neutral gas cools
before it transits off the face of the star.  Starting from where the
mixing layer is well-developed (say the lower red dashed line in
Figure \ref{mixing_layer}), hot neutral gas is advected off the
projected stellar limb in a time
\begin{equation}
t_{\rm adv} \sim 2 R_p / v_\ast \sim 2 \times 10^3 \, {\rm s} \,.
\end{equation}
By comparison, the cooling time is of order 
\begin{equation} \label{tcool}
t_{\rm cool} \sim \frac{1}{n_c^+ \sigma v_{\rm rel}} \sim 500 \left( \frac{2 \times 10^6 \, {\rm cm}^{-3}}{n_c^+} \right) \left( \frac{ 10^{-16} \, {\rm cm}^2}{\sigma} \right) \left( \frac{100 \,{\rm km/s}}{v_{\rm rel}} \right) \, {\rm s}
\end{equation}
where $n_c^+$ is the density of cold ionized hydrogen in the mixing
layer, $v_{\rm rel}$ is the relative speed between hot and cold
hydrogen, and $\sigma$ is the H-H$^+$ cross section for slowing down
fast hydrogen, here taken to be the ``viscosity'' cross section
calculated by \citet{Schultz2008}.\footnote{For slowing down fast H in
  a sea of cold H$^+$, there may also be a contribution to $\sigma$
  from ``momentum transfer'' in ``elastic''
  (non-charge-exchange) collisions. This contribution increases
  $\sigma$ over the viscosity cross section by only $\sim$30\%; compare
  Figures 6 and 7 of \citet{Schultz2008}.}  Our estimate of $t_{\rm
  cool}$ in (\ref{tcool}) neglects cooling by neutral-neutral
collisions, but we estimate the correction to be small, as $n_c^0$ is
lower than $n_c^+$ by a factor of $1/(1-f_p^+) \sim 5$, and the
cross section for H-H collisions is generally not greater than for
H-H$^+$ collisions (A.~Glassgold 2012, personal communication; see
also \citealt{Swenson1985}; note that \citet{Ekenback2010} take the
relevant H-H cross section to be $10^{-17}$ cm$^2$ but do not provide a
reference).

That $t_{\rm cool} \sim t_{\rm adv}$ indicates our simulated
column densities of hot neutral hydrogen may be too large, but
hopefully not by factors of more than a few. Keeping more careful
track of the velocity distributions---and excitation states---of
neutral hydrogen in the mixing layer would not only improve upon
our calculations of Lyman-$\alpha$ absorption,
but would also bear upon the recent
detection of Balmer H$\alpha$ absorption in the hot Jupiters HD 209458b and
HD 189733b (\citealt{jensen2012}).

\revised{
\item {\it Magnetic fields.} Insofar as our results depend on
  Kelvin-Helmholtz mixing, our neglect of magnetic fields is worrisome
  because magnetic tension can suppress the Kelvin-Helmholtz
  instability (\citealt{frank1996}). For numerical simulations
  of magnetized planetary winds interacting with magnetized stellar winds,
  see \citet{cohenetal11a,cohenetal11b}. These magnetohydrodynamic
  simulations
  can track how planetary plasma is shaped by Lorentz forces, but
  as yet do not resolve how the planetary wind mixes and exchanges charge with
  the stellar wind.
}

\item {\it Dependence of Ly-$\alpha$ absorption $A$ on the planetary
    wind density $n_p$.} In the same vein as item (ii), we found
  empirically that $A\propto n_p^{1/2}$, and argued that this result
  arose from the Kelvin-Helmholtz growth timescale.  Ekenb\"ack et
  al.~(2010) found a much weaker dependence: increasing $n_p$ by a
  factor of 100 only increases $A$ in their models by a factor of
  $\sim$2 at -100 km/s and even less at positive velocities---see
  their Figures 8 and 9.  The true dependence of $A$ on $n_p$ remains
  unclear.

\item {\it Rotational effects and gravity.}  There are a few
  order-unity geometrical corrections that our study is missing. Our
  standard stellar wind velocity of $v_\ast = 130$ km/s is comparable
  to the planet's orbital velocity of $v_{\rm orb} = 150$ km/s, so
  that in reality the stellar wind strikes the planet at an angle of
  roughly 45 deg. The Coriolis force will also deflect the planetary
  wind by an order-unity angle after a dynamical time of $r/v_{\rm
    orb} \sim 5\times 10^4$ s, by which time the wind will have
  travelled $\sim$$5 R_p$ from the planet.  These geometrical effects are
  potentially observable---see, e.g., 
\revised{
\citet{Schneiter:2007ju} and \citet{Ehrenreich:2008hj} for modeling
of HD 209458b,} and
\citet{Rappaport2012} for a
  real-life example of a transit light curve that reflects the
  trailing comet-tail-like shape of the occulting cloud. However, these
geometrical effects seem unlikely
  to change the basic order of magnitude of the absorption $A \sim 10$\%
  that we have calculated.

  We have also neglected planetary gravity, stellar tidal gravity, and
  the centrifugal force, all of which can change the planetary wind
  velocity. But
  this omission seems minor, since we have drawn our input planetary
  wind velocities from calculations that do account for
  such forces (M09), at least along the substellar ray.  According to
  Figure 9 of M09, the planetary wind accelerates from $v_p
  \approx 10$ km/s at a planetocentric distance $d = 4R_p$, to $v_p
  \approx 30$ km/s at $d = 10R_p$. This range of velocities and
  corresponding distances overlap reasonably well with the range of
  velocities and distances characterizing our simulations.

\item {\it Hydrodynamic approximations for the stellar and planetary winds.} We have
  not formally justified our use of the hydrodynamic equations to
  describe the wind-wind interaction. The problem is that
  the collisional mean free path in the stellar wind is
  much longer than the lengthscales of the flow:
  $\lambda_{\rm Coulomb, \ast} = 1/(n_\ast\sigma_{\rm Coulomb}) \sim
  10^{13} (10^4 \, {\rm cm}^{-3}/n_\ast) (10^{-17} \,{\rm cm}^{2} /
  \sigma_{\rm Coulomb}) \, {\rm cm}$, where $\sigma_{\rm Coulomb} \sim
  10^{-17} (T_\ast / 10^6 \, {\rm K})^{-2} \, {\rm cm}^2$ is the cross
  section for protons scattering off protons. That the Solar wind
  is collisionless and does not necessarily admit a one-fluid treatment
  is well-known.

  Nevertheless, it is perhaps just as well-known that
  Parker's (\citeyear{Parker58}, \citeyear{Parker63})
  use of the fluid equations to describe
  the collisionless Solar wind is surprisingly accurate, capturing the
  leading-order features of the actual Solar wind. The role of Coulomb
  collisions in relaxing the velocity distribution functions of
  protons and electrons is fulfilled instead by plasma instabilities
  and wave-particle interactions---see, e.g., reviews of Solar wind
  physics by \citet{Marsch03} and \citet{Marsch06}. The
  gross properties of collisionless shocks can still be modeled
  with the hydrodynamic equations insofar as those properties depend only on the
  macroscopic physics of mass, momentum, and energy conservation, and
  not on microphysics (e.g., \citealt{Shu92}).

  Note that the planetary wind is fully collisional because
  of its higher density and lower temperature, and modeling it as a single fluid
  appears justified: $\lambda_{{\rm Coulomb}, p} \sim 10^7
  (10^6 \, {\rm cm}^{-3} / n_p) (T_p/10^4 \,{\rm K})^2 \, {\rm cm}$, which is
  smaller than any other length scale in the problem.

\item {\it Non-Maxwellian behavior of the stellar proton velocity
    distribution.} Lyman-$\alpha$ absorption at the
  redshifted velocity of +100 km/s arises from charge-exchanged
  neutral hydrogen at the assumed stellar wind temperature of $10^6$
  K.  We have assumed a Maxwellian distribution function for hydrogen
  in the stellar wind, and have ignored non-Maxwellian features that
  have been observed in the actual Solar wind, including high-energy
  tails and temperature anisotropies. Accounting for non-Maxwellian
  behavior may introduce order-unity corrections to our results for
  the absorption. For the more polar fast Solar wind, proton
  temperatures parallel to and perpendicular to the Solar wind
  magnetic field differ by factors of a few at heliocentric distances
  of 5--10 Solar radii \citep{McKenzie97}.  For the more equatorial slow
  Solar wind---which our simulations are modeled after---temperature
  anisotropies are more muted (\citealt{Marsch03}, page 391).

\item {\it Stellar radiation pressure.} Stellar Lyman-$\alpha$ photons
  can radiatively accelerate neutral hydrogen away from the star
  (e.g., \citealt{Vidal-Madjar2003}; M09).  Both
  the planetary wind, and the charge-exchanged stellar wind in the mixing
  layer, are subject to a radiation pressure force that exceeds the
  force of stellar gravity by a factor $\beta$ on the order of
  unity.

  Radiative repulsion of the charge-exchanged stellar wind in the
  mixing layer may not be observable, because once hot neutral
  hydrogen is created in the mixing layer, it is advected off the
  projected limb of the star before radiation pressure can produce a
  significant velocity: $\delta v_{\rm rad} \sim GM_\ast / r^2 \times
  \beta \times t_{\rm adv} \sim 6 \beta$ km/s, which does not exceed
  the hot neutral hydrogen's thermal velocity of $\sim$100 km/s.

  What about radiative acceleration of the planetary wind? The
  travel time of the planetary wind from the planet to the mixing
  layer is $\sim$$10 R_p / v_p \sim 10^5$ s, long enough for neutral
  hydrogen to attain radiative blow-out velocities in excess of 100
  km/s.  However, the amount of hydrogen that suffers radiative
  blow-out is limited to the column that presents optical depth unity
  to Lyman-$\alpha$ photons. This column is $1/\sigma_{\rm line-ctr}
  \sim 2 \times 10^{13} (T_p/10^4 \, {\rm K})^{1/2}$ cm$^{-2}$, and is
  much smaller than the typical column in the planetary wind, which is
  $(1-f_p^+) n_p R_p \sim 10^{16}$ cm$^{-2}$.  
Thus the bulk of the
  planetary wind is shielded from radiative blow-out, and our neglect
  of radiation pressure appears safe.  
\revised{ 
  Note that \citet[][]{LecavelierdesEtangs:2012jq} find that
  radiation pressure cannot explain the largest blueshifted velocities
  observed for HD 189733b; like us, they favor charge exchange.
}

\end{enumerate}

\section*{\small{Acknowledgments}}
\small{ This project was initiated during the 2011 International
  Summer Institute for Modeling in Astrophysics (ISIMA) program hosted
  by the Kavli Institute for Astronomy and Astrophysics (KIAA) at
  Beijing University. We thank Pascale Garaud and Doug Lin for
  organizing this stimulating workshop, and Edouard Audit, Stuart
  Bale, Andrew Cumming, Sebastien Fromang, Pascale Garaud, Astrid
  Lamberts, Eliot Quataert, and Jim Stone for helpful exchanges. We
  are indebted to Al Glassgold for his guidance in helping us
  understand H-H and H-H$^+$ collisions, and an anonymous referee for
  a report that improved the presentation of this paper.  ISIMA is
  funded by the American and Chinese National Science Foundations, the
  Center for the Origin, Dynamics and Evolution of Planets and the
  Center for Theoretical Astrophysics at the University of California
  at Santa Cruz, the Silk Road Project, and the Excellence Cluster
  Universe at TU Munich. Support for EC was provided by NASA through a
  Hubble Space Telescope Theory grant. This work was granted access to
  the HPC resources of [CCRT/CINES/IDRIS] under the allocations
  c2011042204 and c2012042204 made by GENCI (Grand Equipement National
  de Calcul Intensif).}

%
%

\bibliographystyle{aa}
\bibliography{bib_paper.bib}






\end {document}